\documentclass[review]{fcs}

\usepackage{multirow}
\usepackage{float}
\usepackage{wrapfig}
\usepackage{verbatim}
\usepackage{algorithmicx,algorithm}
\usepackage{stfloats}
\usepackage{makecell}
\usepackage{lipsum}
\usepackage{cuted}
\usepackage{caption}
\usepackage{subfigure}
\usepackage{xcolor}
\usepackage{pifont}
\usepackage{enumitem}
\usepackage{amsmath}
\usepackage{amsthm}
\setenumerate[1]{itemsep=0pt,partopsep=0pt,parsep=\parskip,topsep=5pt}
\setitemize[1]{itemsep=0pt,partopsep=0pt,parsep=\parskip,topsep=5pt}
\setdescription{itemsep=0pt,partopsep=0pt,parsep=\parskip,topsep=5pt}

\usepackage{color}

\usepackage{lscape}
\usepackage{lineno}
\usepackage{fancyhdr}
\fancypagestyle{mypagestyle}{
\fancyhead{}
}

\title{Quality Control in Open-Ended Crowdsourcing: A Survey}
\author[1]{Lei Chai}
\author*[1,2]{Hailong Sun}
\author[1]{Jing Zhang}
\address[1]{State Key Laboratory of Complex \& Critical Software Environment (CCSE), Beihang University, Beijing 100191, China}
\address[2]{Hangzhou Innovation
Institute of Beihang University, Hangzhou 310056, China}
\fcssetup{
  received       = {month dd, yyyy},
  accepted       = {month dd, yyyy},
  corr-email     = {xxx@xxx.xxx},
}
\begin{abstract}
Crowdsourcing provides a flexible approach for leveraging human intelligence to solve large-scale problems, gaining widespread acceptance in domains like intelligent information processing, social decision-making, and crowd ideation. However, the uncertainty of participants significantly compromises the answer quality, sparking substantial research interest. Existing surveys predominantly concentrate on quality control in Boolean tasks, which are generally formulated as simple label classification, ranking, or numerical prediction. Ubiquitous open-ended tasks like question-answering, translation, and semantic segmentation have not been sufficiently discussed. These tasks usually have large to infinite answer spaces and non-unique acceptable answers, posing significant challenges for quality assurance. This survey focuses on quality control methods applicable to open-ended tasks in crowdsourcing. We propose a two-tiered framework to categorize related works. The first tier introduces a holistic view of the quality model, encompassing key aspects like task, worker, answer, and system. The second tier refines the classification into more detailed categories, including `quality dimensions’, `evaluation metrics’, and `design decisions’, providing insights into the internal structures of the quality control framework in each aspect. We thoroughly investigate how these quality control methods are implemented in state-of-the-art works and discuss key challenges and potential future research directions.
\end{abstract}
\keywords{Crowdsourcing, Open-Ended Task, Quality Control}

\begin{document}

\section{Introduction}
\label{sec:intro}

Posting questions online and receiving answers from other users become much easier for internet users in the era of modern web technologies. Open-ended crowdsourcing has been widely adopted for collecting data~\cite{guo2015mobile} for artificial intelligence models that require large-scale labeled datasets such as text translation~\cite{li2019dataset,li2020crowdsourced,braylan2020modeling}, image segmentation~\cite{cheng2001color} and object recognition~\cite{song2020c,maninis2018deep,sofiiuk2020f,jang2019interactive}. Open-ended crowdsourcing also enables the solution of complex tasks that are not possible yet to solve through automatic methods. For instance, workers in open-ended crowdsoucing tasks collaborate to write novels~\cite{kim2017mechanical}, solve mysteries~\cite{li2018crowdia} and create taxonomies~\cite{chilton2013cascade}.

In addition to the potential in solving complex tasks, there are several reasons why open-ended crowdsourcing is of great significance. First, it allows for the acquisition of fine-grained information. Compared to Boolean crowdsourcing~\cite{de2015reliable}, 
open-ended crowdsourcing is an efficient way to introduce more detailed information to the computing system. 
For instance, image segmentation, a typical open-ended task, is crucial for scene understanding. Rather than assigning a single label to an entire image, the segmentation task essentially requires fine-grained annotations from crowd workers to identify the location and segmentation boundaries of sub-targets within an image through pixel-level labeling. Second, it is cost-effective. For example, given a set of 3D-point cloud data, employing a few experienced experts to annotate the target objects is costly and inefficient, while posting open-ended crowdsourcing tasks on crowdsourcing platforms is more effective in terms of time and cost. 


As a result, both industry and academia have utilized open-ended crowdsourcing to tackle complex tasks due to its capabilities. For instance, in terms of advancing computer vision technology, crowd workers are employed to provide fine-grained image information by annotating the category of interrelated pixel points in the image, such as object boundaries. A notable accomplishment in this field is VisualGenome~\cite{krishna2017visual}, which has become a critical benchmark for related research and applications (e.g., image segmentation~\cite{lee2018aggregating} and Autopilot system~\cite{rzadca2020autopilot}). In scientific research, citizen participants are generally asked to provide open-ended answers to scientific questions. For instance, in disease diagnosis tasks, diagnostic reports with different concerns and diverse language styles may be collaboratively generated by crowd workers~\cite{han2021exploring}. Consequently, citizen science projects have addressed numerous challenges in areas such as astronomical observation~\cite{fortson2012galaxy}, protein folding~\cite{jumper2021highly}, biological population analysis~\cite{khare2016crowdsourcing} and disease diagnosis~\cite{han2021exploring}. For knowledge production, over 80 million Wikipedia users have collaborated to write nearly 200 million entries~\footnote{https://www.wikipedia.org/} in various expressions, writing styles and even languages. Therefore, open-ended crowdsourcing is emerging as an indispensable approach for accomplishing large-scale complex tasks with low cost and high efficiency.

Consequently, the core issue in open-ended crowdsourcing is how to enhance the quality of responses through statistical modeling or mechanism design. Given that the participants for open-ended crowdsourcing are dynamically recruited from the Internet, crowd workers have uncertain capabilities~\cite{wang2021teaching,gadiraju2019crowd} and motivations~\cite{chen2022adversarial, rechkemmer2020motivating}. Their motivation, ability level, and time commitment vary greatly and may be influenced by various factors like time, environment, and psychological state, thus posing serious quality problems in answers. 





\subsection{Representative Open-Ended Crowdsourcing Tasks}
Considering the prevalence and diversity of open-ended crowdsourcing tasks in real-world scenarios, we categorize the tasks of open-ended crowdsourcing into three types: `intelligent information processing', `crowd social decision-making' and `crowd ideation'. 
We will present some typical open-ended crowdsourcing tasks, which we use as examples in this paper.

\textbf{1) Intelligent information processing.} In this paper, we primarily concentrate on information-processing tasks that can be used for supervised learning. Open-ended crowdsourcing tasks for intelligent information processing can be bifurcated into two categories: 1. direct tasks that employ crowdsourced answers as solutions, such as optical character recognition~\cite{mithe2013optical}, animal audio classification~\cite{nanni2020data}; 2. indirect tasks that are associated with knowledge acquisition: these tasks indirectly provide solutions by extracting and organizing knowledge, such as knowledge mapping~\cite{wexler2001and} and knowledge base construction~\cite{shin2015incremental}. Simple crowdsourcing tasks like numerical prediction, binary classification and ranking provide labeled data for the training and deployment of AI models, thereby accelerating the implementation and application of AI algorithms in areas like face recognition~\cite{zhao2003face}, intelligent customer service bots~\cite{chen2019antprophet} and pedestrian recognition~\cite{gray2008viewpoint}. As AI technology continues to evolve, emerging applications such as autonomous driving~\cite{levinson2011towards}, smart cities~\cite{yaqoob2017enabling}, and metaverse~\cite{lee2021all} have created new demand for labeled data. Some representative open-ended tasks of intelligent information processing include:
\begin{itemize}
    \item Image segmentation. Image segmentation~\cite{cheng2001color} is generally regarded as a pixel-level classification task that focuses on subdividing an image into multiple sub-regions (or collections of pixels) by identifying objects and boundaries (such as lines and curves) within the image.
    \item Open-ended question answering. The objective of open-ended question answering is to collect responses to specific questions from the crowd~\cite{chai2022error}. The peculiarity of such tasks lies in the fact that the answer can be subjective and heavily influenced by the worker's capability. 
    \item 3) Translation. Translation tasks aim to request crowd workers to provide translations in the target language~\cite{li2019dataset}, based on the sentences in the source language.
\end{itemize}   

    
\textbf{2) Crowd social decision-making.} Crowd social decision-making is an effective approach to solving specific social issues, such as drawing on the opinions and preferences of the crowds~\cite{li2022context}, finding social influencers~\cite{arous2020opencrowd} and acquiring specific knowledge~\cite{han2021find}. 
    
\textbf{3) Crowd ideation.} The main challenge in crowd ideation is to assemble distributed groups of workers with minimal mutual interaction~\cite{schmidt2016using} while generating creative content, like stories~\cite{kim2016storia,huang2020heteroglossia},  plans~\cite{deng2021multistep}, and summaries~\cite{verroios2014context,wang2018exploring,zhang2017wikum}. 

To avoid redundancy, we have selected one representative work for each type of task (such as image segmentation and audio transcription) and summarized them in Table~\ref{tab:open-ended tasks}. Given that the scale of the answer significantly impacts the design of answer modeling and answer aggregation methods, we have organized and summarized the size of answer space (number of acceptable answers) for open-ended tasks and divided it into three categories: \textit{countable}, \textit{large but countable} and \textit{large and uncountable}.

We found that simple Boolean crowdsourcing is not suitable for most open-ended tasks. For instance, an emergency event cannot be adequately described with finite category labels or continuous values. Therefore, exploring modeling and optimization methods for open-ended crowdsourcing is both crucial and challenging.

\subsection{Quality Control in Open-Ended Crowdsourcing}

Most previous research on optimizing crowdsourcing has concentrated on Boolean tasks. However, according to AMT log data~\cite{marcus2015crowdsourced} and crowdsourcing industry user surveys~\cite{difallah2015dynamics}, the demand for open-ended tasks in the crowdsourcing market has surpassed traditional simple crowdsourcing tasks and continues to rise. 

Quality control methods(like D\&S~\cite{dawid1979maximum}, ZenCrowd~\cite{demartini2012zencrowd}, BCC~\cite{kim2012bayesian} and LFC~\cite{raykar2010learning}) for Boolean crowdsourcing have been widely explored over the past years. However, open-ended tasks have been overlooked due to their inherent characteristics like complex task structure, large to infinite answer space and non-unique acceptable answers.
For instance, the authors of ImageNet~\cite{deng2009imagenet} argue that `computers still perform poorly on cognitive tasks such as image description and question-answering', and propose an image dataset `Visual Genome~\cite{krishna2017visual}' with complex annotations including attribute annotation, semantic segmentation, image description, and question-answering. While both ImageNet~\cite{deng2009imagenet} and VisualGenome~\cite{krishna2017visual} aim to identify the image content, annotation tasks of VisualGenome impose more fine-grained requirements: for example, indicating the relationship between the objects in the image; describing the content of the image in open-ended text; designing question-answer pairs to explore the semantic information in the image. In the case of VisualGenome~\cite{krishna2017visual}, crowd workers may need to provide complex statements that come from infinite answer space. 
In summary, there are general differences between Boolean crowdsourcing and open-ended crowdsourcing.

Consider that `quality' has diverse connotations in different research areas such as data management, artificial intelligence and software engineering. In this paper, we adopt the definition of quality proposed by Crosby~\cite{crosby1979quality} as a guide to identify the target and dimensions of the quality control issue, where `conformance to requirements' is regarded as the core requirement. Therefore, `quality control' primarily focuses on how to produce high-quality answers to meet the needs of the requester. Broader implications of quality control, such as security and reliability, are beyond the scope of this paper. The problem of quality control in open-ended crowdsourcing is defined as follows: 

\begin{definition} 
    (Quality Control in Open-ended Crowdsourcing).\\
    Given a set of open-ended tasks $T$ published by a set of requesters, a group of publicly recruited crowd workers $W$ capable of providing answers, and a set of answers $A$ provided by the crowd workers. Quality control in open-ended crowdsourcing refers to the improvement of the quality of answer set $A$ through a collection of statistical modeling and design methods, denoted as $\Theta$, so that the final answer set $A’$ could meet the requesters' quality requirements. The functions of these methods include but are not limited to optimizing the design of crowdsourcing tasks, enhancing the capabilities of crowd workers, and eliminating noise in crowdsourced answers.
\end{definition}


Specifically, the quality control issue in open-ended crowdsourcing focuses on optimizing methods for tasks with no exact evaluation metrics and unique ground truth, implying that there may be more than one acceptable answer. The answers for these tasks generally come from a large to infinite answer space, making it uncommon for different workers to provide identical answers to the same task. In addition, these tasks may consist of a set of interdependent subtasks, so that context information and worker collaboration may significantly contribute to the improvement of answer quality. 

\begin{table*}[t]
    \centering
    \caption{A brief review of representative open-ended tasks in crowdsourcing. \textit{IIP} refers to Intelligent Information Processing, \textit{CSDm} denotes the task of Crowd Social Decision-making, \textit{CI} refers to Crowd Ideation, \textit{Size} refers to the size of the acceptable answer, \textit{large-co} denotes large but countable, \textit{large-un} refers to large and uncountable}
    \begin{tabular}{l|c|c|l|l|l}
         \textbf{Name} & \textbf{Data Type} & \textbf{Task Type} & \textbf{Open-ended Task} & \textbf{Answer} &  \textbf{Size}\\
         \hline
         f-BRS~\cite{sofiiuk2020f} & image & IIP & semantic segmentation & clicks on the positive and negative area & large-un\\
         \hline
         PDMRR~\cite{braylan2021aggregating} & image & IIP & human body annotation  & keypoints on the human skeleton & large-co\\
         \hline
         CrowdTruth~\cite{inel2014crowdtruth} & image & IIP & image description  & event description according to the image & large-un\\
         \hline
         LATTE~\cite{wang2019latte} & image & IIP & LiDAR  & clicks on the center of the target object & large-un\\
         \hline
         Lean~\cite{branson2017lean} & image & IIP &  object selection & 2-D bounding box on the target object & large-co\\
         \hline
         Cascade~\cite{chilton2013cascade} & image & CSDm & taxonomy creation  & suggested categories of the colors & large-co\\
         \hline
         Gadiraju et al.~\cite{gadiraju2019crowd} & image & IIP & image transcription  & decipher characters from a given image & large-un\\
         \hline
         Trainbot~\cite{abbas2020trainbot} & text & IIP/CSDm & question-answering  & answers to open-ended questions  & large-un\\
         \hline
         BAU~\cite{braylan2020modeling} & text & IIP & sequence annotation  & the text span of the target content  & large-un\\
         \hline
         MAS~\cite{braylan2020modeling} & text & IIP & sentence translation  & translations of the initial sentence  & large-un\\
         \hline
         Nguyen et al.~\cite{nguyen2017argument} & text & CSDm & opinion aggregation  & textual ideas  & large-un\\
         \hline
         Context trees~\cite{verroios2014context} & text & CSDm & importance rating  & importance value of event fragments  & countable\\
         \hline
         Kaur et al.~\cite{kaur2018creating} & text & CI & plan creation  & textual plans  & large-un\\
         \hline
         CrowDEA~\cite{baba2020crowdea} & text & CI & ideas ranking  & ideas and opinions  & large-un\\
         \hline
         Trainbot~\cite{abbas2020trainbot} & text & CSDm/IIP & conversation  & multiple rounds answers & large-un\\
         \hline
         Nguyen et al.~\cite{hung2017computing} & text & CSDm & information finding  & middle-names of personalities & large-un\\
         \hline
         GONCALVES et al.~\cite{goncalves2017eliciting} & text & IIP/CSDm & knowledge eliciting  & pairs of options and criteria & large-un\\
         \hline
         CrowdTruth~\cite{inel2014crowdtruth} & text & IIP/CSDm & NER  & relation between entities & countable\\
         \hline
         QAMR~\cite{michael2017crowdsourcing} & text & IIP/CSDm & semantic mining  & question-answer pairs & large-un\\
         \hline
         Heteroglossia~\cite{huang2020heteroglossia} & text & CI & story ideation  & suggestion of story ideas & large-un\\
         \hline
         TAS~\cite{schmitz2018online} & text & CSDm & news writing  & pieces of events & large-un\\
         \hline
         FFV~\cite{tran2015crowdsourcing} & text & IIP/CSDm & sentence correction  & revised sentence & large-un\\
         \hline
         VoxDIY~\cite{pavlichenko2021crowdspeech} & audio & IIP/CSDm & audio transcription  & textual content of speech & large-un\\
         \hline
         Lipping et al.~\cite{lipping2019crowdsourcing} & audio & IIP/CSDm & audio captioning  & textual summary of speech content & large-un\\
         \hline
         Kaspar et al.~\cite{kaspar2018crowd} & video & IIP & video segmentation  & scribbles on the video frame & large-un\\
         \hline
         CrowdTruth~\cite{inel2014crowdtruth} & video & IIP/CSDm & video description  & textual descriptions of video content & large-un\\
         \hline
         C-Reference~\cite{song2020c} & video & IIP & object selection  & 2-D bounding box & large-co\\
         \hline
         Event anchor~\cite{deng2021eventanchor} & video & IIP/CSDm & context of object and event  & 2-D bounding box & large-un\\
         \hline
         Popup~\cite{song2019popup} & video & IIP & position identification  & dimension lines on objects & large-co\\
         \hline
         PDMRR~\cite{braylan2021aggregating} & numerical & IIP/CSDm & id ranking  & numerical sequence & countable\\
         \hline
    \end{tabular}
    \label{tab:open-ended tasks}
\end{table*}

\subsection{Challenges to Quality Control in Open-ended Crowdsourcing}

\label{sec:chall}

Given that the quality control methods in Boolean crowdsourcing generally make certain assumptions~\cite{paun2019proceedings} that do not exist in open-ended crowdsourcing: 1) the candidate answers are independent of each other, 2) answers come from a finite answer space, and 3) there are exact evaluation metrics, so a unique true answer exists for a specific task. As a result, most existing quality control methods are only applicable to Boolean crowdsourcing tasks, like rating~\cite{li2014confidence} and classification~\cite{aydin2014crowdsourcing}, and thus are not suitable for open-ended crowdsourcing tasks. To intuitively introduce the relationship between Boolean and open-ended crowdsourcing, we present the algorithm flow of crowdsourcing quality control, along with the similarities and differences between them, as shown in Figure~\ref{fig:fra}. It is worth noting that the boxes in blue denote challenges that do not exist in Boolean crowdsourcing, while boxes in gray indicate challenges that commonly exist in both Boolean crowdsourcing and open-ended crowdsourcing but have distinct differences, like 'dynamic and multi-dimensional worker abilities'.


Without loss of generality, the challenges of quality control in open-ended crowdsourcing lie in the following aspects:

\textbf{1) Dynamic and Multidimensional Worker Abilities.} 
Accurately estimating worker abilities is crucial for assigning the task to the appropriate worker, thereby improving the quality of answers. Unlike Boolean crowdsourcing tasks, the answer quality of specialized open-ended crowdsourcing tasks (such as question answering, story creation and translation) is usually influenced by the knowledge background of crowd workers and has a certain degree of subjectivity. Therefore, crowd workers' abilities for open-ended crowdsourcing tasks are multidimensional. For instance, it is important to evaluate whether the crowd workers have the background knowledge to complete the complex open-ended task, and how well they have mastered the knowledge, rather than representing the worker's ability with a number ranging from 0 to 1. Moreover, changes in workers' abilities over the course of the task are not negligible: workers become more adept at completing specific open-ended tasks as the number of completed tasks increases. 
    
\textbf{2) Complex Task Structure.} Unlike simple micro-tasks, open-ended tasks usually consist of a set of interdependent subtasks organized by complex workflows. For instance, in an intelligence analysis task, crowd workers usually have access to only a portion of the information. Therefore, decomposing the complete task into interrelated subtasks allows crowd workers to access context information, which is important for producing comprehensive, high-quality intelligence analysis results.

\textbf{3) Multiple Acceptable Answers.} Many open-ended tasks like opinion gathering, story writing, and event reporting does not have unique acceptable answers. The variability of workers' knowledge, skills and subjective biases leads to the diversity of answers. Beyond that, for open-ended crowdsourcing tasks, most crowd workers usually struggle to provide complete answers, leading to diversity and limitation of the answers. Since the final output is generally a consistent result obtained by aggregating participants' answers, the diversity of crowd contributions makes it more challenging to derive a `ground truth' with high confidence.

\textbf{4) Difficulty in Evaluation and Aggregation.} In Boolean crowdsourcing, probabilistic graphical model (PGM)~\cite{koller2009probabilistic}-based algorithms (like D\&S~\cite{dawid1979maximum} and its extensions~\cite{venanzi2014community}) are widely
applied for answer evaluation and aggregation. 
However, such methods cannot be directly applied to open-ended crowdsourcing due to several reasons. First, diverse contributions: given that there may be several acceptable answers, it becomes challenging to accurately assess the contribution of each answer, thus posing challenges for answer aggregation. Second, uncertainty in answer representation: for automated evaluation, answers are generally mapped into numerical values or dense vectors. Different embedding methods (such as `Universal Sentence Encoder'~\cite{cer2018universal} and `BERT'~\cite{devlin2018bert}) may focus on different aspects of a textual answer’s semantic information, derive different similarity relationships between answers and the truth, and thus impact the result of answer evaluation and aggregation. Third, significant differences across open-ended tasks make it challenging to find uniform evaluation and embedding methods. 
    

In summary, issue $1$ indicates the difficulty of worker estimation and task assignment; issue $2$ leads to challenges in task analysis and workflow design; issues $3$ and $4$ point out problems in answer modeling and aggregation.

\subsection{Related Survey}
Numerous surveys on crowdsourcing have been published in recent years. Some provide a general overview of crowdsourcing~\cite{vaughan2017making}, while others focus on specific aspects such as 1) certain steps in the crowdsourcing process (e.g., worker organization~\cite{wang2021examination} and task design~\cite{dellermann2021future}); 2) specific types of tasks (e.g., writing tasks~\cite{feldman2021we} and medical image analysis~\cite{orting2019survey}), and 3) specific research fields (e.g., computing vision~\cite{kovashka2016crowdsourcing} and natural language processing~\cite{paun2019proceedings}).

Recently, with the increased focus on quality control issues, surveys on crowdsourced quality control have begun to emerge. For instance, Li et al.~\cite{zheng2017truth} summarized 17 truth inference algorithms in crowdsourcing to assist users in understanding existing truth inference algorithms and conducting detailed comparison experiments, providing readers with guidance for selecting appropriate methods for different tasks. Jin et al.~\cite{jin2020technical} offered a unified taxonomy of the crowdsourcing quality control methods from the perspective of mechanism design and statistical methods. Daniel et al.~\cite{daniel2018quality} discussed this topic from the viewpoint of quality attributes, assessment techniques and assurance actions. However, most of the quality control methods discussed in these surveys are only suitable for Boolean crowdsourcing tasks. Quality control methods for open-ended crowdsourcing tasks urgently need to be reviewed and discussed.

\subsection{Contributions}
This survey aims to elaborate on quality control issues in open-ended crowdsourcing. It seeks to assist requesters, workers, and developers of crowdsourcing applications in understanding the roles of quality control methods in the process of open-ended crowdsourcing and how to enhance answer quality through appropriate methods. Specifically, it provides the following contributions:
\begin{itemize}
    \item A definition of open-ended crowdsourcing and the scope of quality control methods.
    \item An introduction to representative open-ended crowdsourcing tasks and the challenges in quality control.
    \item A two-tiered framework of all the related works, including a holistic view of the quality model encompassing key aspects: task, worker, answer and context, along with finer-grained categories by ‘quality dimensions’, ‘evaluation metrics’, and ‘design decisions’ for each aspect.
    \item A systematic review of all related works, which are organized by the two-tiered framework.  
    \item An in-depth discussion of research status and identification of current research challenges along with respective opportunities for future research, especially in the era of Large Language Models (LLMs).
\end{itemize}
\section{Quality Control in Open-Ended Crowdsourcing}
\label{sec:quality}

\begin{figure*}
\centering\includegraphics[width=0.90\textwidth]{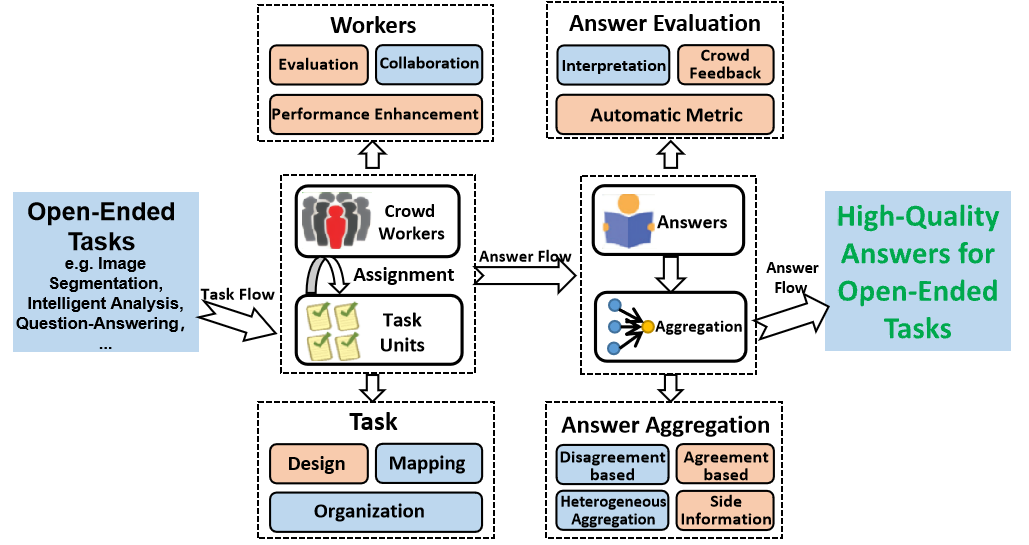}
\caption{The Execution Process and Algorithm Framework of Crowdsourcing Quality Control: Round boxes in blue do not exist in Boolean crowdsourcing, while boxes in orange represent distinct differences between Boolean crowdsourcing and Open-ended crowdsourcing.}
\label{fig:fra}
\end{figure*}

Before we introduce the quality control framework in detail, it is important to identify the execution process and key aspects involved in open-ended crowdsourcing. According to the task execution process adopted by representative crowdsourcing platforms (such as AMT, Scale AI, and Appen), we have summarized the organization and execution process of crowdsourcing tasks, and highlighted the key parts and crucial topics in quality control issues, as shown in Figure~\ref{fig:fra}.

Generally, open-ended crowdsourcing tasks are executed by the flow of tasks and answers. 

Task flow refers to the key steps in the organization and flow of tasks in crowdsourcing. Initially, macro open-ended tasks are decomposed into sub-tasks that are suitable for crowd workers to complete. Subsequently, the micro-tasks are assigned to appropriate workers (e.g., with the matched knowledge background or task experience). After that, micro-tasks are solved by crowd workers according to the task requirements. 

Answer flow refers to the key process of answer generation and optimization. Firstly, crowd workers provide the initial answers according to the requirements of micro-tasks. Then, the answer reliability is evaluated based on relevant parameters, such as answer error and worker reliability. After that, the optimized answer (also known as aggregated answer or estimated truth) is derived based on the answer and its reliability parameters. 

\subsection{Key Aspects of Quality Control in Open-ended Crowdsourcing} 
Current research on open-ended crowdsourcing quality control is dedicated to exploring the relationship between key aspects and answer quality~\cite{jin2020technical}. Therefore we specify four critical aspects of open-ended crowdsourcing related to the issue of quality control before we conduct the systematic review on this topic. 

\textbf{1) Task.} In crowdsourcing, tasks are usually decomposed into easy-to-complete micro-tasks to facilitate crowd workers in providing high-quality answers. Generally, a micro-task is the smallest unit that is distributed to workers~\cite{wu2021task} in crowdsourcing.
Open-ended crowdsourcing tasks can be categorized into `single task' and `contextual task' based on the task structure. A single task mainly asks for independent open-ended answers from workers. For example, annotating an object with a 3D frame in a 3D point cloud image. In contrast, a contextual task can usually be considered as a set of interdependent sub-tasks, which may require considering the context information of other sub-tasks when completing the task. For example, captioning video content based on the content of certain frames. Furthermore, the task may be subjective, where there are multiple acceptable answers.

\textbf{2) Crowd Worker.} Crowd worker (hereafter termed as worker) denotes the users involved in answering questions with certain motivation, typically recruited dynamically from internet platforms in crowdsourcing. As discussed in  section~\ref{sec:chall}, workers need to be modeled dynamically and multidimensionally in open-ended crowdsourcing. Multidimensional modeling and analysis of workers aim to assign the most appropriate task to each worker to produce high-quality answers. 
    
\textbf{3) Answer (also known as response).} The responses to the task provided by workers are referred to as answers. In quality control, the final answer is generally derived by aggregating all the answers to the same task, and referred to as the `estimated truth'. Answers are generally collected from a large to infinite solutions space in open-ended crowdsourcing.
    
    
\textbf{4) System.} Quality control in open-ended crowdsourcing is a system-level problem that is typically discussed within the context of multiple quality dimensions. In this paper, it mainly consists of cross-aspect quality control methods and workflow design. 
We found that multiple quality dimensions are generally jointly considered while modeling and optimizing the answer quality. 
Workflow greatly aids in identifying the execution steps and organizing crowd workers in the process of solving open-ended tasks, especially for `contextual tasks'. 
\subsection{Holistic View of the Quality Model}

\begin{figure}[t]
\centering\includegraphics[width=\linewidth,height=\linewidth]{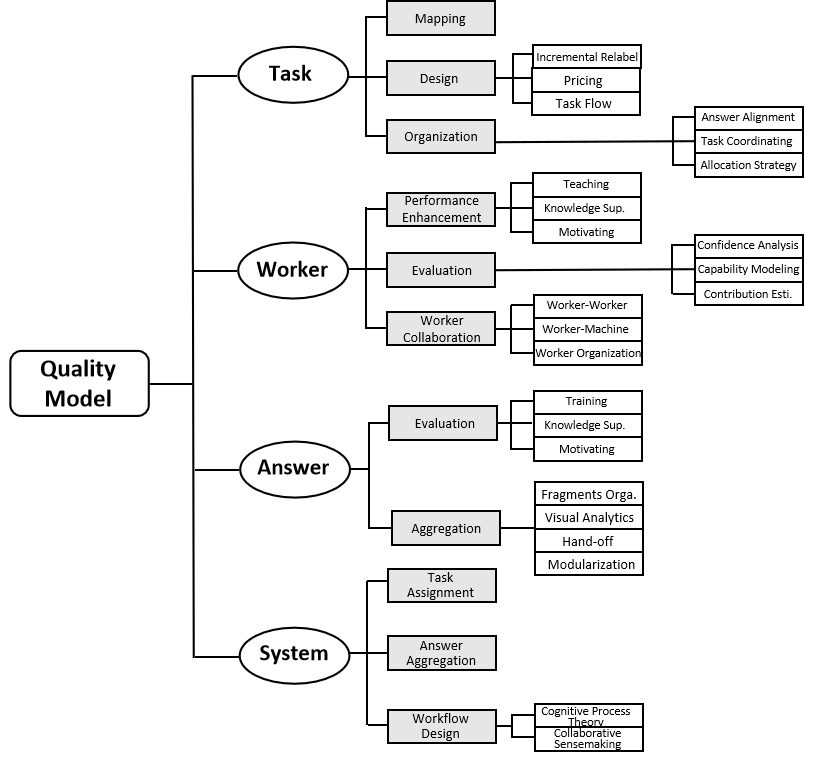}
\caption{Quality model in open-ended crowdsourcing. Each ellipse node represents a specific aspect with dimensions (boxes in gray) and design decisions (boxes without fill).}
\label{fig:model}
\end{figure}

Most existing open-ended crowdsourcing quality control studies aim to improve the answer quality through statistical modeling or design methods on one or more specific aspects (including task, worker and answer ) introduced above.
\label{sec:model}

We propose a comprehensive view of quality control methods in open-ended crowdsourcing, as depicted in Figure~\ref{fig:model}. The quality model introduces the aspects, their related dimensions, and the design decisions considered in past research works for quality control methods designed in open-ended crowdsourcing. It is worth noting that a dimension denotes an important branch of the related aspect that has been jointly considered by numerous crowdsourcing quality control research works, a design decision signifies a specific class of research ideas for addressing quality control problems. 

\subsection{Taxonomy in Each Aspect}
\begin{figure*}[t]
\centering\includegraphics[width=0.7\textwidth]{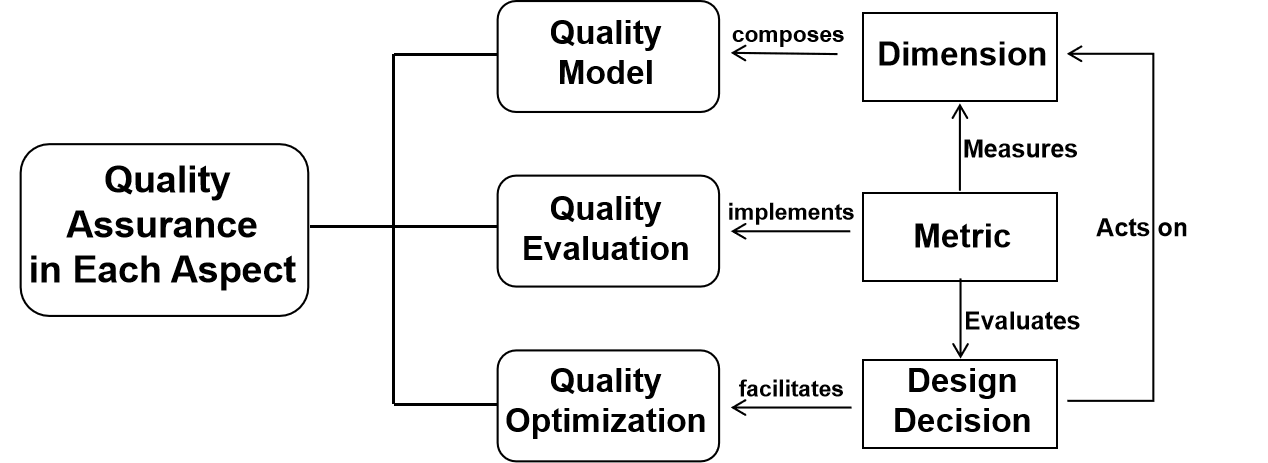}
\caption{The internal structures of the quality control framework in each aspect.}
\label{fig:inter}
\end{figure*}

Considering that the current fragmented literature lacks an analysis of the fine-grained categories of research works, we developed a taxonomy shown in Figure~\ref{fig:inter} to
study and analyze what is the scope of the quality control issue and how quality is dealt with in each aspect. The proposed taxonomy aims to highlight the hierarchical relationship between the related quality control works, which greatly assists readers in identifying the attributes, evaluation metrics and design decisions when developing quality control methods. Specifically, the taxonomy consists of three dimensions: 

\textbf{1) Quality model.} in each aspect refers to a collection of quality dimensions, which are used to describe key attributes that may impact the quality of crowdsourcing tasks. Quality dimensions are typically an abstract representation of a category of methods and issues related to an aspect, such as task mapping and task organization. For each aspect, understanding the quality dimensions can help readers find an appropriate entry point when they explore to optimize crowdsourcing quality.

\textbf{2) Quality evaluation.} aims to measure the impact of specific quality dimensions and optimization methods on the quality of crowdsourcing results. Accurate evaluation results are crucial for designing effective quality control methods~\cite{hu2020quality}. Quality assessment is generally implemented by a series of evaluation metrics. For automatic metrics,  `annotation distance'~\cite{braylan2020modeling} and `answer similarity'~\cite{li2018crowdia} are often considered general metrics for answer assessment; other automated evaluation metrics, like `GLEU'~\cite{mutton2007gleu}, and `Intersect Over Union (IoU)', are generally adopted for textual and image data. Beyond that, subjective evaluation is also a common evaluation method in open-ended crowdsourcing, such as peer grading~\cite{zhu2014reviewing}, expert feedback~\cite{deng2021eventanchor} and human-AI collaborative approach~\cite{mesbah2023hybrideval}.
    
\textbf{3) Quality optimization.} is the final step in open-ended crowdsourcing quality control, aimed at improving the final quality of crowd-sourced answers. Specifically, researchers optimize the quality of each stage of the crowdsourcing task by making design decisions that act on the quality dimensions defined by the quality model, such as designing incentive mechanisms to improve worker motivation and designing task decomposition and reconstruction methods to reduce task difficulty.


We have organized the remainder of the research works according to the proposed quality model and taxonomy. Each section corresponds to one aspect in the quality model, with `quality dimensions', `evaluation metrics' and `design decisions' reviewed.  

\subsection{Literature Selection}

We compiled a list of conferences and journals that publish research works on crowdsourcing and related topics to identify the papers to be considered in this survey. The conferences considered include AAAI, IJCAI, WWW, KDD, SIGIR, HCOMP, ICML, NIPS, CHI, CSCW, VLDB, CIKM, WSDM, UBICOMP, UIST, CVPR, EMNLP, ACL, ICDE, ICIP, DASFAA and NAACL. The journals include ACM CSUR, VLDBJ, JAIR, AIJ, IEEE TKDE, IJCV, TPAMI, IEEE TNNLS, TVCG, ACM TOCHI, ACM TOCS, ACM TWEB, ACM TIST and Information Sciences. We conducted a search for papers published between 2012 and 2023 with the following keywords in either the title, abstract or keywords: Complex Crowdsourcing, Open-ended Crowdsourcing, Crowd transcription , Crowd Translation, Crowd Image description and Complex annotation. The search returned 419 papers and the authors read the title and abstract of each paper. Papers were excluded from further consideration if they focused on simple crowdsourcing tasks (like binary classification) or if they did not involve the topic of quality control, which resulted in a list of 147 papers for further consideration. Additional papers considered are derived from the authors’ prior knowledge.

\section{Task Model}
\label{sec:task}
Tasks, which are usually posted by requesters on public internet platforms, serve as the starting point of crowdsourcing. In open-ended crowdsourcing, tasks can be categorized into two types: objective tasks and subjective tasks. Objective tasks typically have clear evaluation criteria and a limited answer space. These tasks are usually more complex compared to Boolean crowdsourcing, requiring more human cognitive effort and sophisticated task design to yield high-quality answers. 

\subsection{Quality Dimensions}

A significant number of researchers in crowdsourcing quality control focus on the topic of task modeling and organizing. The method of task construction and implementation directly influences the quality of the answers provided by crowd workers, and significantly impacts the quality of open-ended crowdsourcing. We identify the following quality dimensions:

\textbf{1) Task design.} is the starting point in task modeling, and the results determine the content and interface of the task, which has a direct impact on the difficulty of crowd workers in completing tasks. Therefore, task design is a crucial step in task modeling. Considering that open-ended crowdsourcing tasks usually have complex task structures and annotation content, the process task design mainly includes steps such as task decomposition and reconstruction~\cite{xu2014voyant}, task pricing~\cite{singer2013pricing}, communication mechanisms design~\cite{salehi2017communicating}, and pipeline/workflow design~\cite{tran2015crowdsourcing}. 
    
\textbf{2) User interface.} is where crowd workers complete tasks and submit answers. The interface can either be an HTML integrated into the crowdsourcing platform (such as the AMT task dashboard\footnote{https://www.mturk.com/}) or a separate annotation application (such as Labelme\footnote{http://labelme.csail.mit.edu/Release3.0/}). These user interfaces commonly integrate annotation tools and are connected by a series of processes. Therefore, the quality of the user interface determines how easily crowd workers can complete tasks~\cite{abbas2020trainbot}.
    
\textbf{3) Incentives.} refer to the extrinsic (e.g., monetary reward) or intrinsic rewards (e.g., worker prestige) that requesters provide to crowd workers through the platform after completing crowdsourcing tasks. The design of incentives directly affects the enthusiasm of crowdsourcing workers in completing tasks, and therefore, has a direct impact on the quality of crowdsourced answers. 

There are also quality dimensions related to task modeling that may necessitate interaction between aspects. For instance, the quality dimension `task allocation' typically requires an estimation of the worker’s ability and knowledge background, in addition to modeling the content and difficulty of the task itself. We will provide a comprehensive introduction to these quality dimensions with cross-aspect interactions in Section~\ref{sec:context}.
%



\subsection{Quality Evaluation}
The primary purpose of quality evaluation for task modeling is to evaluate the quality of task organization and design, with the aim of guiding the design of crowdsourcing tasks and thereby enhancing the quality of crowdsourced answers. Considering that numerous researchers explore measuring the quality of crowdsourcing task design through the quality of crowdsourced answers, we categorize the quality evaluation methods of task modeling into two types: direct and indirect. Moreover, considering the prevalent existence of subjective tasks in open-ended crowdsourcing and the subjectivity characteristic of task quality, the evaluation of task design should be considered by both objective indicators (such as numbers of operations~\cite{wang2019latte}) and subjective metrics (such as crowd feedback). 

\textbf{1) Objective Evaluation.} Number of automated metrics are utilized to objectively evaluate the quality of task modeling. On one hand, some researchers strive to determine whether the designed crowdsourcing tasks are easy to complete by analyzing the cost required to complete crowdsourcing tasks, such as monetary expenses~\cite{tran2015crowdsourcing, doroudi2016toward}, time~\cite{cychnerski2021segmentation, benenson2019large}, number of operations~\cite{wang2019latte}, completion rate~\cite{schmitz2018online}. On the other hand, statistical metrics about quality, such as average error rate~\cite{rechkemmer2021exploring, song2019popup}, number of operations for target accuracy~\cite{sofiiuk2020f}, and engagement probability~\cite{allahbakhsh2015task} are frequently used to indirectly evaluate the quality of task design. Furthermore, some automated indicators that reflect the participation of crowd workers are also used to measure the quality of task modeling indirectly. For instance, Rechkemmer et al.~\cite{rechkemmer2021exploring} measure the extent to which the design of task instructions can enhance the ability of workers to complete tasks through the learning gain indicator. Additionally, mental demand scores~\cite{salehi2017communicating} are utilized to identify the difficulty of the designed tasks by measuring the mental cost of crowd workers in completing specific tasks.
 
\textbf{2) Subjective Evaluation.} Given that the answer to `the quality of task modeling' often varies among users, many researchers explore to subjectively evaluate the quality of task modeling by collecting feedback from crowd workers. For example, some researchers directly measure the quality of task design by asking workers to provide comments~\cite{huang2020heteroglossia, xu2014voyant} on the dimensions of usability, helpfulness, clarity, and conciseness for task and interface designs. In addition, rating is also a common method to help workers provide subjective evaluations. Evaluation methods such as peer rating~\cite{zhu2014reviewing}, expert rating~\cite{bragg2018sprout, abbas2020trainbot}, and self-rating~\cite{doroudi2016toward} are generally adopted for the assessment of task design and answer quality.



\subsection{Quality Control Methods}

\begin{figure}[t]
\centering\includegraphics[width=9cm, height=8cm]{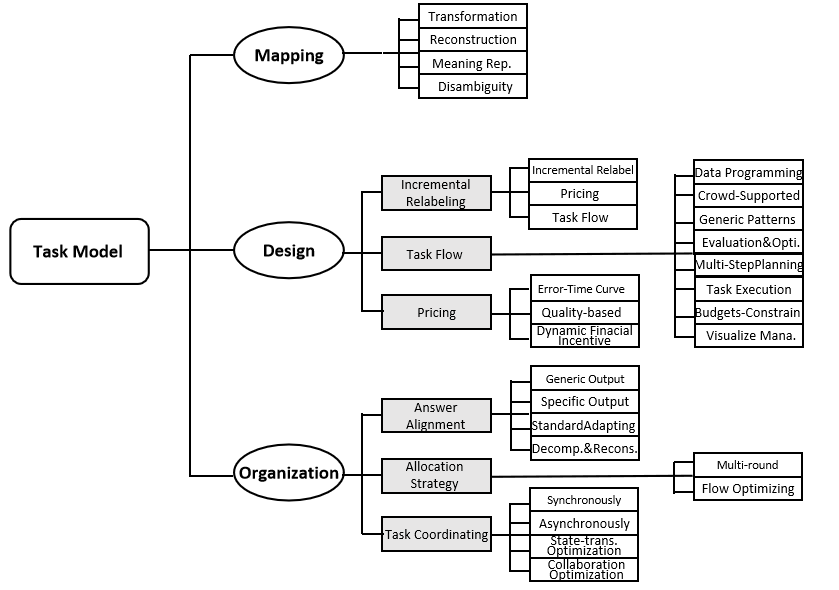}
\caption{Task model in open-ended crowdsourcing. Each ellipse node represents an important branch with design decisions(boxes in gray) and specific research works (boxes without fill).}
\label{fig:task1}
\end{figure}

According to the quality dimensions and evaluation metrics introduced above, we introduce the quality control method of task modeling from the following three aspects:

\subsubsection{Task Mapping}
\label{sec:simp}
Asking novice crowds to provide high-quality answers directly is difficult and costly for open-ended tasks due to the huge cognitive burden and workload. An intuitive approach to task simplification is to transform complex crowdsourcing tasks into simple annotating tasks. For instance, complex scene recognition tasks are transformed into 2D annotation tasks~\cite{song2020c} and numerical annotation tasks~\cite{braylan2021aggregating} with crowd-machine hybrid approaches. Additionally, some works explore transforming crowdsourcing tasks into binary annotation tasks through multi-step planning~\cite{deng2021multistep} and disambiguation~\cite{dumitrache2018capturing}. Furthermore, meaning representation~\cite{michael2017crowdsourcing} and reconstruction~\cite{braylan2021aggregating} are also adopted to simplify complex open-ended crowdsourcing tasks.

\subsubsection{Workflow Design}
\label{sec:design}
The workflow design is one of the most critical parts in task modeling, and the logic of task flow directly influences the efficiency of task execution and the efficiency of crowd workers. Specifically, the ideas of workflow design may originate from the crowd~\cite{bragg2018sprout,biemann2013creating}, generic workflow patterns~\cite{gadiraju2019crowd} and budget constraints~\cite{tran2015crowdsourcing}. The tasks may be implemented according to the results of automatic workflow optimization~\cite{boer2016pplib,de2017efficiently,bhuiyan2020investigating}, multi-step planning~\cite{deng2021multistep}, instant execution~\cite{suzuki2018crowdsheet}, data programming~\cite{dunnmon2020cross} and visualized management~\cite{kittur2012crowdweaver}. Moreover, pricing is also an important factor in task and workflow design given that crowdsourcing is a profit-oriented computing paradigm. Compared to a quality-based pricing strategy~\cite{wang2013quality}, the proposed methods of the error-time curve~\cite{cheng2015measuring} and dynamic financial incentive~\cite{yin2013effects} are more helpful in adapting to changes in worker capabilities as complex tasks are performed.


\subsubsection{Task Organization}
\label{sec:organ}


Appropriate methods of task organization significantly impact the overall quality and efficiency of the macro-crowdsourcing task, especially since open-ended tasks typically comprise a set of interdependent subtasks. For instance, in multi-step crowdsourcing tasks like intelligence analysis and event reporting, the sequencing and organization of the steps critically influence the final output. Numerous research works have discussed the organizational relationships between subtasks from both synchronous~\cite{agapie2015crowdsourcing,lasecki2013chorus} and asynchronous~\cite{li2018crowdia,zhu2014reviewing,drapeau2016microtalk} perspectives. Additionally, some researchers consider the problem of task organization as an optimization problem and explore the optimal method of task organization using state machine~\cite{xiong2018smartcrowd} and collaboration optimization~\cite{rahman2019optimized}.
\section{Worker Model}
\label{sec:worker}

In open-ended crowdsourcing, tasks usually possess a complex structure and necessitate certain professional knowledge. Completing such complex tasks requires not only a redundant processing mechanism for individual subtasks but also the collaboration of multiple participants. Therefore, enhancing the quality of open-ended crowdsourcing through worker-centered modeling and optimization method design primarily focus on the following key issues: how to accurately describe multi-dimensional dynamic worker capabilities; how to improve the expertise of crowd workers to obtain high-quality answers; and how to design incentive mechanisms to boost the subjective initiative of workers. 

\begin{figure}[t]
\centering\includegraphics[width=9cm,height=9cm]{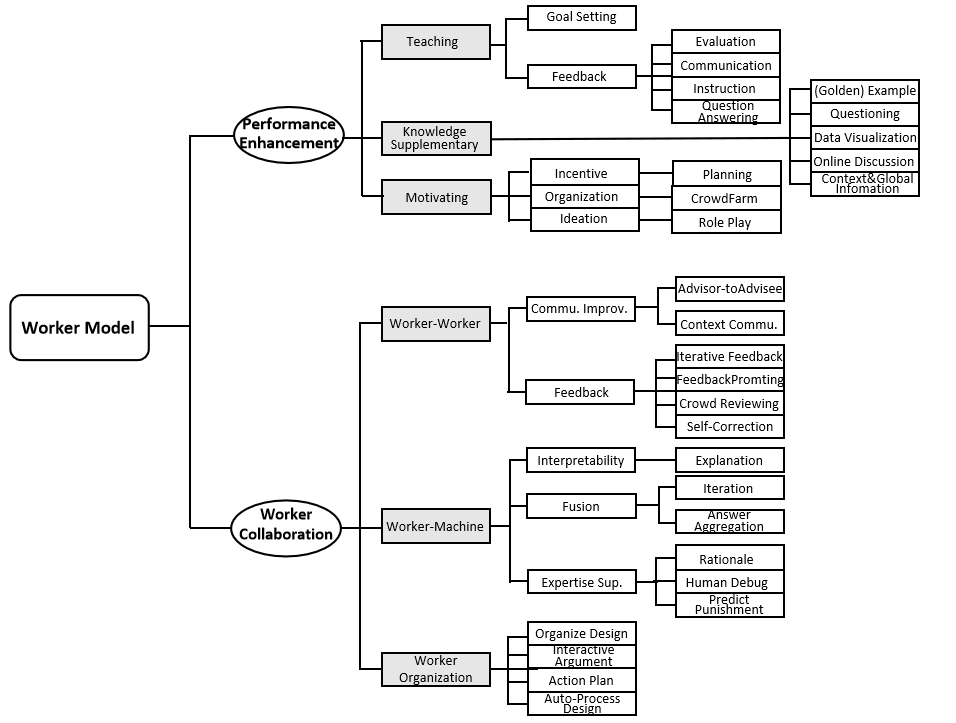}
\caption{Worker model in open-ended crowdsourcing. The ellipse nodes represent important branches with quality attributes (boxes in gray) and specific design decisions (boxes without fill).}
\label{fig:task}
\end{figure}

\subsection{Quality Dimensions}
Modeling and analysis of crowd workers is a critical topic of crowdsourcing quality control. The strategy employed for worker evaluation and selection directly influences the quality of the answers, thereby significantly impacting the completion quality of open-ended crowdsourcing. This discussion encompasses quality control approaches for worker modeling, focusing on expertise modeling, motivating design and contribution analysis.

\textbf{1) Expertise modeling.} Expertise directly determines whether crowd workers can provide correct answers for specific tasks. Workers’ expertise is generally influenced by diverse factors including skills, educational background, task experience, age, gender, etc., and exhibits different characteristics under different motivations and work states. Modeling worker expertise not only helps assign crowdsourcing tasks to suitable workers but can also be used to assess the quality of answers and measure workers’ contributions. In general, expertise modeling helps optimize the quality of open-ended crowdsourcing at multiple stages of the crowdsourcing workflow.
    
\textbf{2) Incentive design.} A worker’s motivation determines the effort they invest in each task and the sincerity of their answers. 
The sincerity of a worker’s response reflects the probability of them providing what they believe to be the correct answer. Workers with malicious intent are more likely to provide incorrect answers. Therefore, designing effective incentive methods can enhance the seriousness and concentration of crowd workers when completing tasks, thereby improving the quality of answers. 
    
\textbf{3) Contribution analyzing.} Unlike Boolean crowdsourcing, estimating the contribution of a crowd worker to completing a specific open-ended task is very challenging. This challenge primarily stems from two aspects: 1) Open-ended crowdsourcing tasks may consist of a series of interrelated subtasks, crowd workers may complete one or more subtasks, thus resulting in a lack of complete information about the task for the crowd workers. 2) Open-ended crowdsourcing may include subjective tasks, and the answers provided by workers could be partially correct or conditionally correct. Consequently, assessing the contribution of crowd workers is crucial for task allocation, response aggregation, and reward determination in open-ended crowdsourcing.

\subsection{Quality Evaluation}
\label{sec:workereva}
Existing research on worker evaluation can be broadly categorized into confidence analysis, capability modeling, and contribution estimation.

Due to the complexity of open-ended tasks, exact matching usually cannot meet the needs of its correctness evaluation. 
As a result, some researchers have proposed to estimate the confidence of workers through partial agreement~\cite{hung2017computing} and behavior analysis~\cite{pei2021quality,gadiraju2019crowd} in open-ended crowdsourcing tasks.
    
\subsubsection{Expertise evaluation} (also referred to as capability modeling or skill modeling) is another significant topic in worker modeling. 
Considering that the state and ability of workers in the process of handling tasks are changing, researchers explored estimating worker capability at a fine-grained~\cite{cui2017complex} and dynamic~\cite{gadiraju2019crowd} level through multidimensional skill modeling. In particular, representative approaches of skill modeling methods include Single-Skill Modeling~\cite{tang2020optimal}, Multi-Skill Modeling~\cite{mavridis2016using}, Skill Ontology Modeling~\cite{maarry2014skill} and Cognitive-based Modeling~\cite{hettiachchi2020crowdcog}.
    
\subsubsection{Contribution estimation} As discussed in section\ref{sec:chall}, the consistent results are difficult to infer due to the diversity of crowd contributions in open-ended crowdsourcing. Existing research works on contribution assessment for open-ended crowdsourcing can be divided into two categories, crowd-based~\cite{aris2019review} methods and answer quality-based methods. The crowd-based methods evaluate the contribution of answers based on feedback from other novice crowds~\cite{zlabinger2020dexa} and domain experts~\cite{xu2014voyant}. Other works mainly focus on assessing the answer contribution with answer similarity~\cite{li2020crowdsourced} or annotation distance~\cite{braylan2020modeling}.

\subsection{Quality Control Methods}
Numerous research topics in worker modeling, such as worker evaluation and teaching, are pertinent to both Boolean and open-ended crowdsourcing. However, the design decisions vary significantly due to the difference in the dimension of worker expertise and worker organization. As a result, we categorize the quality control methods for worker modeling into the following categories: 

\subsubsection{Performance Enhancement}
\label{sec:enh}
Compared to Boolean crowdsourcing tasks, obtaining high-quality answers directly from novice crowd workers for open-ended tasks is usually challenging due to the requirement of substantial cognitive effort and domain-specific knowledge.
Consequently, numerous researchers have explored methods to improve the performance of crowd workers (especially novice crowds) to enhance the quality of answers for open-ended crowdsourcing tasks, including:

\textbf{1) Teaching.} Teaching serves as a method to enable novice workers to tackle complex crowdsourcing tasks. Rechkemmer et al.~\cite{rechkemmer2020motivating,rechkemmer2021exploring} proposed to set different goals (including performance goals, learning goals, and behavioral goals) for worker teaching to significantly improve the workers' ability to handle complex crowdsourcing tasks. Furthermore, researchers have discovered that crowd workers can effectively learn from diverse types of feedback due to humans’ inherent perception and learning abilities, as evidenced by findings from the research area of machine teaching~\cite{wang2021teaching,zhu2015machine}. Therefore, feedback emerges as another crucial factor in enhancing worker performance when dealing with complex open-ended crowdsourcing tasks. A summary of related work indicates that feedback from multiple sources could have a teaching effect on novice crowd workers, such as multi-turn contextual argumentation~\cite{chen2019cicero}, peer communication~\cite{tang2019leveraging}, gated instruction~\cite{liu2016effective}, and automatic conversational interface~\cite{abbas2020trainbot}.
    
\textbf{2) Knowledge Supplementary.} In addition to worker teaching, numerous researchers have proposed auxiliary methods to enhance the level of information exposure of workers. These methods include addressing workers’ task-related questions through online discussion~\cite{mcinnis2018crafting,zhang2017wikum} and question answering~\cite{hettiachchi2021challenge}, displaying golden examples to workers~\cite{doroudi2016toward}, presenting multidimensional task information to workers through data visualization~\cite{willett2012strategies}, exposure to contextual and global information by distributed information synthesis and evaluation~\cite{hahn2016knowledge,verroios2014context}.
    
\textbf{3) Motivating.} For creative open-ended tasks like crowd writing and product design, it is crucial to develop mechanisms and tools that provide on-the-spot assistance when the workers get stuck, in addition to monetary~\cite{yin2013effects} incentives. For instance, Kaur et al.~\cite{kaur2018creating} proposed vocabulary-based planning to motivate workers in complex writing tasks, Wang et al. proposed a reputation management~\cite{wang2021examination}-based approach to organize and motivate workers, Huang et al. proposed an in-situ ideation~\cite{huang2020heteroglossia} method for crowd writing tasks.

\subsubsection{Worker Collaboration}
\label{sec:coll}
Collaboration plays a significant role in completing tasks with complex structures and heavy cognitive load. 
Numerous research works have explored the design of collaborative mechanisms to enable novice crowds to accomplish large-scale open-ended tasks. 
Collaboration methods can be categorized into worker-worker-based and worker-machine-based collaboration methods depending on different collaboration subjects.

\textbf{1) Worker-Worker Collaboration} Worker collaboration helps to integrate diverse knowledge and skills from the Internet to accomplish open-ended crowdsourcing tasks. 
It also facilitates the communication of contextual information between the interdependent sub-tasks. 
Consequently, researchers are exploring ways to enhance the answer quality through worker collaboration. For instance, researchers have explored improving worker communication with the mechanical design of Advisor-to-Advisee~\cite{kaur2018creating} and context communication~\cite{salehi2017communicating}; others have designed feedback mechanisms to achieve worker collaboration through Iteration Feedback~\cite{xu2015classroom}, Feedback Prompting~\cite{huang2017supporting}, Crowd Reviewing~\cite{zhu2014reviewing} and Self-Correction~\cite{kobayashi2018empirical}. 

\textbf{2) Worker-Machine Collaboration} Humans and machines naturally collaborate well~\cite{tsvetkova2017understanding}: computer algorithms methodically solve large-scale problems with well-defined iterative steps, while humans excel at solving problems requiring cognitive and reasoning skills. In summary, three types of worker-machine collaboration approaches have been designed to enhance answer quality in open-ended crowdsourcing. Firstly, worker explanation is used to improve the interpretability~\cite{jayaram2021human} of machine judgment. Additionally, researchers have explored aggregating the two types of answers directly~\cite{russakovsky2015best,branson2017lean,song2020c,gouravajhala2018eureca} and iteratively~\cite{maninis2018deep,benenson2019large,sofiiuk2020f} to improve the answer quality. Moreover, worker intelligence is injected into the intelligent system as expert knowledge through Rationale~\cite{kanchinadam2020rationale}, Debug Information~\cite{yang2019scalpel,nushi2017human} and Predict Punishment~\cite{das2019joint}.

\section{Answer Model}
\label{sec:answer}
Modeling and optimizing answers are crucial for ensuring the quality of collected answers. The key topic "answer aggregation" (also known as "truth inference") is considered a core issue in crowdsourcing quality control and significantly impacts the quality of the final answer.

Unlike Boolean crowdsourcing, open-ended crowdsourcing presents several challenges for answer evaluation: 1) Answers may be partially correct (e.g. in sequence annotation tasks, some words in a sentence are annotated correctly). 2) There may be more than one answer that can be considered correct. 3) Some subjective evaluation tasks (such as preference collection) may have obvious biases and ambiguities in answers.


\begin{figure}[t]
\centering\includegraphics[width=9cm, height=9cm]{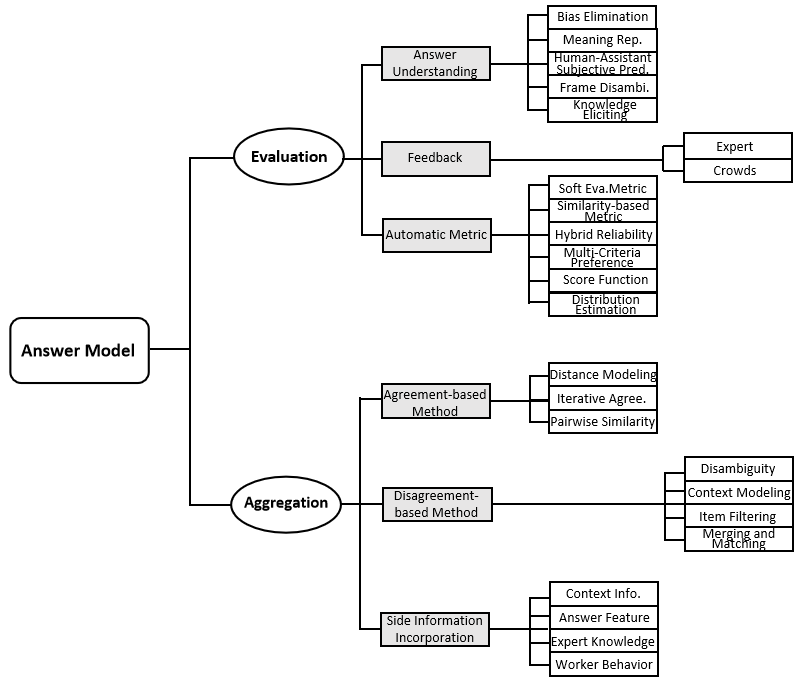}
\caption{Answer model in open-ended crowdsourcing. Each ellipse node represents an important branch with design decisions (boxes in gray) and specific research works (boxes without fill).}
\label{fig:answer}
\end{figure}

\subsection{Quality Dimensions}
Considering that the evaluation results of answers can be used as evidence for deriving the truth in crowdsourcing.
Therefore, a reasonable answer embedding and reliability estimating are the key dimensions in answer modeling.
\subsubsection{Answer embedding} The core of answer embedding is to choose a feature space for the answer, which significantly impacts subsequent answer understanding and processing methods. For example, image annotation tasks generally ask crowd workers to detect objects with bounding boxes, primary content such as the vertex coordinates of the bounding box are generally adopted as the answer to identify the position, size, and interrelationships of the objects. Additionally, the pixel information within the bounding boxes is usually used as the answer embedding, semantic information in the area of bounding boxes becomes the main content conveyed by the answer.
\subsubsection{Answer reliability} The reliability of an answer refers to the likelihood of an answer being correct. Accurately estimating the answer quality plays an indispensable role in inferring ground truth and evaluating worker contributions. Considering the complex task structure and subjective characteristics that are commonly found in open-ended crowdsourcing, the corresponding answer evaluation methods may include both objective metrics and subjective methods (such as peer grading, group evaluation and expert rating). 

\subsection{Quality Evaluation}
\label{sec:ev}

Given the complex structure and various evaluation criteria of answers in open-ended crowdsourcing, numerous researchers have attempted to explore a variety of evaluation methods to assess the quality of the answers as comprehensively as possible. 
`Answer Understanding', `Feedback' and `Automatic Metric' are the representative research topics.

\subsubsection{Answer Understanding} An in-depth understanding of the semantic information in open-ended answers form the basis for accurately evaluating answer quality. On one hand, many researchers explored eliciting knowledge~\cite{goncalves2017eliciting}, and representing the meaning with human rationale~\cite{michael2017crowdsourcing} froms open-ended answers to fully capture the rich knowledge and semantic information contained. On the other hand, some researchers have tried to enhance the quality of open-ended answers by eliminating the subjectivity~\cite{das2019joint}, bias~\cite{simpson2015language,barbosa2019rehumanized} and ambiguity~\cite{dumitrache2018capturing} prevalent in open-ended answers to clarify the meaning.

\subsubsection{Feedback} Feedback serves as another method to evaluate answer quality, taking into account the cognitive differences between participants on a specific task and effectively avoids quality problems due to subjectivity and bias. For instance, DEXA~\cite{zlabinger2020dexa} provides expert feedback by incorporating dynamic examples from experts to crowd workers; Xu et al.~\cite{xu2014voyant} aim to improve the quality of visual design by prompting crowd workers to generate structured feedback; Zhu et al.~\cite{zhu2014reviewing} explore hiring crowd workers to perform answer reviewing to provide feedback.

\subsubsection{Automatic Metric} The use of automatic metrics is the most convenient way to assess the answer quality in Boolean crowdsourcing. Several automatic metrics have been proposed for evaluating the relationship between open-ended answers. For example, $GLEU$~\cite{mutton2007gleu} and $BLEU$~\cite{papineni2002bleu} are commonly used to evaluate the semantic similarity between sentences; $IOU$~\cite{yu2016unitbox} is frequently used to evaluate the accuracy of the image bounding box. However, these automatic metrics are usually related to specific application scenarios and answer types. Therefore, a series of automatic metrics are proposed to evaluate variety types of open-ended answers, like Soft Evaluation Metric~\cite{uma2021learning}, Similarity-based Metric~\cite{braylan2020modeling, ren2024label}, Hybrid Reliability~\cite{li2020crowdsourced}, Multiple-Criteria Preference~\cite{baba2020crowdea}, Neural Network-based Score Function~\cite{vittayakorn2011quality} and Distribution Estimation~\cite{chung2019efficient}. 

\subsection{Quality Control Method: Answer Aggregation}
\label{sec:ag}
Answer aggregation is the most important topic in crowdsourcing quality control.
As previously discussed, modeling the rich semantic information in open-ended answers is important for the answer aggregation in open-ended crowdsourcing. Therefore, 
numerous researchers consider capturing the agreement between answers through similarity-based evaluation metrics (as discussed in section~\ref{sec:ev}) based on the idea of majority Voting, and design agreement-based answer aggregation algorithms. The agreement may take different forms and require different modeling approaches for different tasks. For example, Li et al. propose aggregating pairwise preference with pairwise similarity~\cite{li2021label,li2022context}; Braylan et al.~\cite{braylan2020modeling} explore selecting the best answer with the minimum error which is estimated with annotation distance; Li et al. evaluate the answer reliability dynamically according to iterative agreement estimation~\cite{li2019dataset,li2020crowdsourced}.

Additionally, Dumitrache et al.~\cite{uma2021learning} have found that disagreement is prevalent in open-ended crowdsourcing, which may arise from workers, tasks or annotation schemes. Specifically, ambiguous or unclear~\cite{basile2021we} points in a task may lead to disagreement between answers. 
To address this challenge, Uma et al. propose a disagreement-based computational framework~\cite{uma2021learning,inel2014crowdtruth} to train high-quality models directly from answers with disagreement; Braylan et al. eliminate disagreement in answers by the idea of merging and matching~\cite{braylan2021aggregating}; Timmermans et al. mitigate the semantic gap between answers provided by crowd workers with context modeling~\cite{timmermans2016exploiting}; Klebanov et al. reduce the disagreement among answers by removing highly ambiguous samples through item filtering~\cite{klebanov2014difficult}.

Moreover, in order to integrating fine-grained information about tasks and workers into the answer aggregation process~\cite{zhang2023attribute}, some researchers have tried to integrate specific knowledge by incorporating context information~\cite{lan2019learning,li2022context} to improve the quality of answers. For complex annotation tasks that require domain knowledge (e.g., medical image annotation). Zlabinger et al.~\cite{zlabinger2020dexa} proposed to incorporate expert knowledge to improve the quality of answers provided by novice crowds. 
In addition, some researchers have explored improving the quality of answer aggregation by integrating answer features, such as pixel proximity~\cite{yang2022incorporating} and text features~\cite{simpson2015language}.

\section{System}

\label{sec:context}

Quality dimensions, evaluation approaches and quality control methods of different aspects have been separately introduced in previous sections to help readers gain a holistic view of the quality control issue for open-ended crowdsourcing. However, the issue of quality control in open-ended crowdsourcing is a system-level problem that is typically discussed within the context of multiple quality aspects.
Multiple quality dimensions are generally considered jointly while modeling and optimizing the answer quality. For instance, estimating the reliability of answers usually requires a coordinated consideration of quality dimensions relevant to diverse aspects, such as task difficulty and worker expertise. 

In this section, we will discuss how existing works improve the quality of open-ended crowdsourcing by exploring the interaction of aspects (especially the joint consideration of different quality dimensions) and designing joint optimization methods, in conjunction with the three core tasks in the crowdsourcing execution process: `task allocation', `answer aggregation' and `workflow design'.


\subsection{Task Assignment: Jointly consideration of task design and worker expertise}
Task assignment~\cite{hettiachchi2022survey}, a critical topic of quality control in open-ended crowdsourcing, which is generally discussed in the context of 'task modeling' and 'worker expertise'. The problem of task assignment is generally defined as ~\cite{hettiachchi2022survey}: given a set of questions $\textit{Q}$ = { $q_{1}$, $q_{2}$, ..., $q_{k}$} for a specific task \textit{t} and a set of workers $\textit{W}$ = {$w_{1}$, $w_{2}$, ..., $w_{m}$} where $\lvert|\textit{Q}\rvert|$ = $\textit{k}$ and $\lvert|\textit{W}\rvert|$ = $\textit{m}$. The process of task allocation aims to assign crowdsourcing tasks to the most suitable workers, which may significantly affect the answer quality and completion efficiency.

For Boolean crowdsourcing, parameters such as task categories, task difficulty, worker expertise, worker population and worker practices are generally inferred for the matching of tasks and workers. 
Given the complex cooperative relationship between crowd workers and the dependencies between subtasks in open-ended crowdsourcing, more fine-grained worker evaluation and task modeling are required for task assignment. For example, Cui et al.~\cite{cui2017complex} proposed a PNRN-based multi-round allocation strategy that can dynamically select the allocation strategy and predict the trend of a worker's reputation. Additionally, He et al.~\cite{he2018task} modeled the complex collaborative crowdsourcing task assignment problem as a combinatorial optimization problem based on the maximum flow and computed the optimal solution to task assignment with a Slide-Container Queue (SCP). Beyond that, similar to Boolean crowdsourcing, the utilization of historical worker data~\cite{mo2013cross}, answer distribution~\cite{khan2017crowddqs, fan2015icrowd},
gold questions test~\cite{pavlichenko2021crowdspeech} and behavioral data~\cite{gadiraju2019crowd} are also generally adopted in open-ended crowdsourcing for better worker estimation and task assignment.

\subsection{Answer Aggregation: Jointly consideration of task difficulty, worker expertise and answer quality}
As introduced in Section~\ref{sec:answer}, answer aggregation is the most widely discussed topic in crowdsourcing quality control, which generally considers the joint optimizing of 'task difficulty', 'worker expertise' and 'answer quality'. The problem of answer aggregation is generally defined as given worker's answer $V$, infer the truth $v_{i}^*$ of each task $t_{i}$ $\in$ $T$.

In Boolean crowdsourcing, the difficulty of a task is frequently modeled with a real number~\cite{ma2015faitcrowd}, which is considered highly relevant to the answer quality.
For example, Whitehill et al.~\cite{whitehill2009whose} suggest that the relationship between task difficulty, worker expertise and answer quality can be depicted by the following equation:
\begin{equation}
    Pr(v_{i}^{w}=v_{i}^{*} \mid d_{i},c^{w})=\frac{1}{1+e^{-\frac{c^{w}}{d_{i}}}},
\end{equation}
where $c^{w}$ $\in$ (0,+ $\infty$) denotes the worker ability, $d_{i}$ $\in$ (0, + $\infty$) denotes the task difficulty $t_{i}$. A higher $d_{i}$ represents the higher task difficulty. It can be inferred that for a fixed worker ability $c^{w}$, an easier task (with a lower $d_{i}$) will lead to a higher probability that a worker correctly answers the task. Furthermore, numerous researchers explore modeling the task information with K-dimensional vectors~\cite{blei2003latent,fan2015icrowd,welinder2011multidimensional,zhao2011comparing}, which is helpful in deriving the level of task difficulty.

As discussed in section~\ref{sec:ag}, most existing methods of answer aggregation (such as PGM and optimization-based methods) cannot be directly applied to open-ended crowdsourcing tasks due to the difficulty in measuring dynamic worker expertise and answer reliability. Researchers have investigated incorporating fine-grained information into the process of truth inference, aiming to accurately model the relationships among task difficulty, worker capability, and the quality of answers. A detailed presentation is provided in Section~\ref{sec:ag}, and will not be reiterated in this section. 


\subsection{Workflow Design: Jointly consideration of task design, worker collaboration and human mental model}

\begin{figure}[t]
\centering\includegraphics[width=9cm, height=9cm]{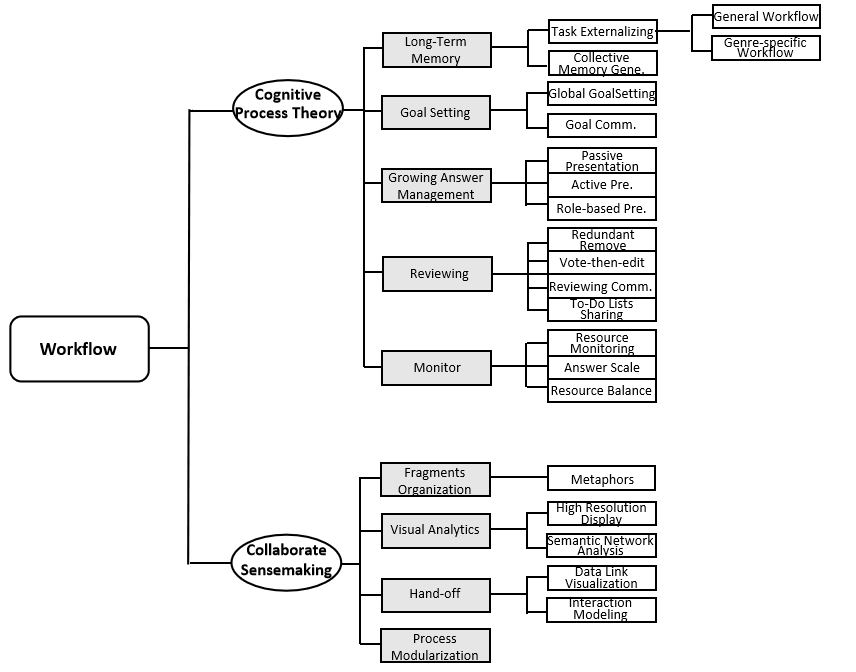}
\caption{Key Theory for Workflow Design in open-ended crowdsourcing. Each ellipse node represents an important branch with design decisions(boxes in gray) and specific research works (boxes without fill).}
\label{fig:workflow}
\end{figure}

Workflow, considered as the dominant infrastructure in crowdsourcing today, is dedicated to decomposing complex macro-tasks into small independent tasks and coordinating worker resources. As discussed in sections~\ref{sec:task} and section~\ref{sec:worker}, numerous researchers propose improving the level of worker collaboration through the workflow design of interactive argumentation~\cite{chen2019cicero} or action planning~\cite{kaur2018creating}; other parts of researchers explore improving the efficiency of task organization with the strategy of multi-round allocation~\cite{cui2017complex} and maximum-flow optimization~\cite{he2018task}. In general, the methods of workflow design in existing works are usually closely related to the task type, and few works discuss the underlying theory. 




\textbf{1) Cognitive Process Theory of Writing}
The theory of `cognitive process' is usually considered as the fundamental theory describing cognitive changes in human writing, which can inspire the design of open-ended text tasks, including  
machine translation, summarization, Q\&A, knowledge production, novel creation, event reporting, etc. 
However, quality control methods used in Boolean crowdsourcing (such as MV, PGM) cannot be applied to crowd-writing tasks directly due to the challenges discussed in Section~\ref{sec:intro}. 
As a result, writing is a representative class of open-ended crowdsourcing tasks, and studying the workflow design theory of crowd-writing can great help explore the research direction of the quality control methods in open-ended crowdsourcing.

The cognitive process model~\cite{flower1981cognitive,feldman2021we} outlines the general guideline of crowd writers, from idea generation to final text completion, as a series of components, including `Topic-related Design', `Worker Identification', `Long-Term Memory', `Goal Setting', `Growing Answer Management', `Reviewing and Monitoring'. Each component of the cognitive process model represents a step of the crowd-writing task, the action and decision in each step directly affect the efficiency and quality of the writing output. Furthermore, Kraut et al.~\cite{kraut1992task} found that the cognitive process model aligns with the process of collaborative writing, which forms the core of workflow design in crowd-writing.

\textbf{2) Collaborative Sensemaking}
As discussed in Section~\ref{sec:coll}, collaboration is a crucial method to integrate diverse knowledge and skills among crowd workers, which is of greatly help to improve the answer quality in open-ended crowdsourcing. However, there is limited discussion on the theoretical foundations for collaborative workflow design, which are vital for persuasiveness and interpretability. 

Sensemaking is defined as the process of searching for a representation and encoding data in that representation to answer task-specific questions~\cite{russell1993cost}. Pirolli et al.~\cite{pirolli2005sensemaking} introduce the sensemaking loop as a "broad brush description" of experts' cognitive process of information transformation. Ericsson et al.~\cite{ericsson1996expert} argued that the key to the expert performance of knowledge-driven open-ended tasks (such as intelligence analysis and crowd writing) is to develop domain-specific schemas from long-term memory and patterns around the key aspects of their tasks. Consequently, numerous researchers~\cite{li2018crowdia} have explored the use of sensemaking as the basic theory for designing workflows that enable distributed novice crowds to solve large-scale cognitive tasks. This topic of workflow design is also referred to as `Collaborative Sensemaking'.
For instance, Bradel et al.~\cite{bradel2013large} proposed a co-located collaborative visual analytics tool for use on large, high-resolution vertical displays. Drieger et al.~\cite{drieger2013semantic} proposed a semantic network-based method to support knowledge building, analytical reasoning and explorative analysis.
These proposals have significantly improved crowdsourcing workflow design, with solutions such as biclusters-based semantic edge bundling~\cite{sun2015biset} and knowledge-transfer graphs~\cite{zhao2017supporting}.



In addition, there exist other basis theories that can be used to guide crowdsourcing workflow design, such as gamification theory~\cite{madge2019crowdsourcing,guillaume2016crowdsourcing,urbanek2019learning} and incentive theory~\cite{muldoon2018survey,scekic2015supporting} (including monetary incentives~\cite{yin2013effects}, worker motivating~\cite{kaur2018creating}, etc.). 

\section{Summary and Discussion}
\label{sec:dis}


In this survey, we begin by introducing representative open-ended crowdsourcing tasks in the context of AI applications. We then summarize and compare the dimensions of quality control research on Boolean crowdsourcing and open-ended crowdsourcing tasks, considering four important aspects, including task, worker, answer and workflow. We found that compared to Boolean crowdsourcing, the quality control research of open-ended crowdsourcing necessitates a more detailed discussion. The dimensions of quality control research on open-ended crowdsourcing significantly exceed those of Boolean crowdsourcing tasks (as shown in Figure~\ref{fig:fra}, Boxes in blue did not exist in boolean crowdsourcing, while boxes in gray represent the distinct differences between Boolean crowdsourcing and complex open-ended crowdsourcing). Therefore, conducting a review of quality control methods on open-ended crowdsourcing is both necessary and challenging. 

We propose a two-tiered framework that encompasses all research on open-ended crowdsourcing quality control. The first tier provides a holistic view of quality control methods, based on the execution process and components in crowdsourcing. This enables readers to understand the key aspects and main characteristics of open-ended crowdsourcing. The second tier analyzes the internal structures of quality control modeling in each aspect from the dimensions of quality models, quality assessment, and quality optimization. This helps readers identify the attributes, evaluation metrics, and design decisions when developing a quality control method, as illustrated in Figure~\ref{fig:inter}.

\subsection{Challenges and Outlook}
According to our survey, existing research on open-ended crowdsourcing quality control has several challenges. Firstly, most quality control methods are applicable to single or a few types of tasks, and few studies have been conducted on generic quality control theories and methods. For instance, Lee et al.~\cite{lee2018aggregating} proposed a quality control method for crowd image segmentation. 
However, this quality control method cannot be applied to other open-ended crowdsourcing tasks, like Q\&A annotation, due to the differences in task structure and prediction algorithms. 
To the best of our knowledge, only a few works focus on discussing the cross-task approaches of answer aggregation~\cite{braylan2020modeling,braylan2021aggregating} and answer evaluation~\cite{braylan2022measuring} for open-ended crowdsourcing.

Another important challenge is that human intelligence has not been fully utilized in all steps in the AI cycle, with most research focusing on improving the quality of labeled data. However, according to Yang et al.~\cite{gadiraju2020can}, human intelligence shows great potential to address the robustness and interpretability issues of AI systems, which may significantly improve the quality of AI applications.

Given the research trends and challenges demonstrated in related literature, we believe that future work related to the following topics will have a long-term impact on the research of open-ended crowdsourcing:

\textbf{1) Domain-specific design.} Current crowdsourcing platforms typically assume that requesters have the sufficient domain knowledge to design reasonable crowdsourcing tasks, corresponding annotation tools and quality control methods. This approach can achieve acceptable results when dealing with some Boolean crowdsourcing tasks. However, this approach often fails to produce satisfactory answers when facing complex open-ended tasks in specific domains, such as medical image recognition, complex 3D point cloud annotation, due to the lack of background knowledge or effective annotation tools. Therefore, domain-specific design for typical tasks in specific domains is a key approach to completing open-ended crowdsourcing tasks and obtaining high-quality answers. 
For instance, Appen launched MatrixGo~\footnote{https://www.appen.com.cn/platform-overview/}, a specific data annotation platform that supports a variety of annotation tools for 2D images and 3D point clouds, to cater to the high-quality annotation data requirements of intelligent cockpits and autonomous driving. Additionally, Wang et al. ~\cite{wang2019latte} focused on designing specific quality control algorithms for image segmentation to improve the data quality for autonomous driving systems. 


\textbf{2) General quality control framework.} Considering that open-ended crowdsourcing generally involve tasks of different data types such as images, videos, and text, the quality control methods may vary significantly due to differences in task structure and answer types. Therefore, most proposed quality control methods are only applicable to a specific task or a few similar tasks. To the best of our knowledge, only a few works~\cite{braylan2020modeling,braylan2021aggregating,braylan2022measuring} have begun to explore quality control methods applicable to open-ended tasks in various data types. Considering the myriad of open-ended tasks in the real world, general quality control methods are both challenging and critically important.


\textbf{3) Intelligent workflow design.} Most current research topics attempt to optimize the completion quality of open-ended crowdsourcing from individual steps. In contrast, as the infrastructure of crowdsourcing, the crowdsourcing workflow can often consider the quality issue from a macro perspective, coordinating and optimizing each link of the task (such as demand analysis, answer collection, task assignment, answer aggregation, etc.) to optimize the overall quality and efficiency. On the one hand, the cognitive process model~\cite{flower1981cognitive} and collaborative perception theory~\cite{pirolli2005sensemaking} will serve as theoretical foundations to explore what kind of workflow is more in line with the human mental model. On the other hand, artificial intelligence technology will be fully involved in every link of the crowdsourcing task, bringing deep optimization to the output of open-ended crowdsourcing. For example, in the stage of demand analysis, LLMs-based generative artificial intelligence applications (like gpt-4~\cite{openai2023gpt}) can transform the requester’s colloquial requirements into specific annotation specifications that can be directly implemented in the process. 


\subsection{Open-ended Crowdsourcing in the Era of LLMs}

Large Language Models (LLMs) have taken the world by storm. LLMs are capable of achieving remarkable results in a range of downstream tasks without gathering extensive task-specific labeled data or tuning model parameters, but merely through the instruction of examples. Given the significant reliance of quality control methods on artificial intelligence algorithms in open-ended crowdsourcing, exploring the relationship between LLMs and open-ended crowdsourcing is of great help in clarifying the key issues and research direction of quality control in open-ended crowdsourcing. This topic will be discussed from the following three aspects:

\textbf{1) Optimizing LLMs through crowd annotations.} Researchers of Open AI~\footnote{https://openai.com/} have integrated human feedback into the training and optimization process of large language models (such as InstructGPT~\cite{ouyang2022training}), in order to reduce the dissemination of harmful and false output. This idea has significantly improved the reliability and safety of large language models. During this process, crowd workers serve as evaluators, ranking the responses of the large language model under different conditions to create a comparative dataset. This comparative dataset is used to train a reward model that can predict which responses will receive higher rankings. Finally, the initial large language model is iteratively fine-tuned using reinforcement learning algorithms and the reward model. This approach continues to be adopted in subsequent processes of research and development in LLMs (e.g. GPT-4~\cite{openai2023gpt}, Claude 2~\cite{bai2022training}, PaLM 2~\cite{anil2023palm}, Llama 2~\cite{touvron2023llama}). Although optimization techniques represented by RLHF~\cite{christiano2017deep} have successfully introduced crowd intelligence into computing systems, this paradigm still has obvious limitations. For example, the quality of annotations is determined by the expertise of the annotators, and the results of model optimization will also have corresponding quality problems. Crowdsourcing, as an important paradigm for introducing human feedback into computing systems, is very likely to provide important insights for future quality optimization methods and development processes of LLMs.


\textbf{2) Inspiring the prompt strategies with design decisions in crowdsourcing.} Wu et al.~\cite{wu2022ai} discovered a high degree of similarity between LLM chains and crowdsourcing workflows. They share common motivations and face similar challenges, such as dealing with cascading errors that affect later stages~\cite{kittur2011crowdforge} or synthesizing inconsistent contributions from crowd workers~\cite{bernstein2010soylent}. A series of recent studies have confirmed this phenomenon: some research has adopted strategies inspired by the crowdsourcing workflow for iteratively improving LLMs, such as self-consistency~\cite{wang2022self} self-inquiry~\cite{press2022measuring} and self-reflection~\cite{shinn2023reflexion}. Furthermore, in balancing the quality and cost of data processing using LLMs, Parameswaran et al.~\cite{parameswaran2023revisiting} viewed LLMs as crowdsourcing workers and leveraged ideas from the literature on declarative crowdsourcing. This includes multiple prompting strategies, ensuring internal consistency, and exploring a mix of LLM and non-LLM methods, to make prompt engineering a more principled process and optimize the workflow of LLM-powered data processing.

\textbf{3) Substituting crowd workers with LLMs.} Due to the powerful semantic understanding and language generation capabilities of LLMs, there is a noticeable trend in using LLMs to replace crowd workers in completing open-ended tasks. Specifically, Veselovsky et al.~\cite{veselovsky2023artificial} found that 33–46\% of crowd
workers used LLMs when completing crowdsourcing tasks. Additionally, Wu et al.~\cite{wu2023llms} took LLMs as a simulation of human workers to construct crowdsourcing pipelines for addressing complex tasks, and further discuss the potential of training humans and LLMs with complementary skill sets. Beyond that, He et al.~\cite{he2023annollm} proposed a two-step approach, ‘explain-then-annotate’ to enable LLMs to generate high-quality labels akin to outstanding crowdsourcing annotators by providing them with sufficient guidance and demonstrated examples.

The aforementioned research works indicate that there is a deep interrelationship between crowdsourcing and LLMs, which brings new demand and inspiration to the study of crowdsourcing quality control: 1) There is a clear trend that Large Language Models (LLMs) are beginning to deeply participate in the process of crowdsourcing tasks. It is necessary to explore new quality control mechanisms to clarify which answers are generated by LLMs and to enable researchers to exclude them or eliminate the contained biases before conducting analysis. 2) Human feedback remains a key step in ensuring the usability of LLMs. Appropriate quality control methods can help the model understand human preferences more accurately, ultimately making the output of LLMs better aligned with human needs.





\section*{Acknowledgement}
This work was supported by National Natural Science Foundation of China under Grant Nos. (61932007, 61972013 and 62472017), and partly by Guangxi Collaborative Innovation Center of Multi-source Information Integration and Intelligent Processing.). We would like to express our gratitude to Prof. Jie Yang and Dr. Ines Arous for their invaluable guidance and suggestions throughout this research. Their expertise and insights are very helpful in the completion of this work. 


\bibliographystyle{fcs}
\bibliography{mybib}

\begin{thebibliography}{100}

\bibitem{ma2015faitcrowd}
Ma~F, Li~Y, Li~Q, Qiu M, Gao J, Zhi S, Su~L, Zhao B, Ji~H, Han J.
\newblock Faitcrowd: Fine grained truth discovery for crowdsourced data aggregation.
\newblock In: Proceedings of the 21th ACM SIGKDD international conference on knowledge discovery and data mining.
\newblock 2015,  745--754

\bibitem{whitehill2009whose}
Whitehill J, Wu~T~f, Bergsma J, Movellan J, Ruvolo P.
\newblock Whose vote should count more: Optimal integration of labels from labelers of unknown expertise.
\newblock Advances in neural information processing systems, 2009, 22

\bibitem{blei2003latent}
Blei D~M, Ng~A~Y, Jordan M~I.
\newblock Latent dirichlet allocation.
\newblock the Journal of machine Learning research, 2003, 3: 993--1022

\bibitem{fan2015icrowd}
Fan J, Li~G, Ooi B~C, Tan K~l, Feng J.
\newblock icrowd: An adaptive crowdsourcing framework.
\newblock In: Proceedings of the 2015 ACM SIGMOD International Conference on Management of Data.
\newblock 2015,  1015--1030

\bibitem{welinder2011multidimensional}
Welinder P, Branson S, Perona P, Belongie S.
\newblock The multidimensional wisdom of crowds.
\newblock 2011

\bibitem{zhao2011comparing}
Zhao W~X, Jiang J, Weng J, He~J, Lim E~P, Yan H, Li~X.
\newblock Comparing twitter and traditional media using topic models.
\newblock In: European conference on information retrieval.
\newblock 2011,  338--349

\bibitem{song2020c}
Song J~Y, Chung J~J~Y, Fouhey D~F, Lasecki W~S.
\newblock C-reference: Improving 2d to 3d object pose estimation accuracy via crowdsourced joint object estimation.
\newblock Proceedings of the ACM on Human-Computer Interaction, 2020, 4(CSCW1): 1--28

\bibitem{deng2021eventanchor}
Deng D, Wu~J, Wang J, Wu~Y, Xie X, Zhou Z, Zhang H, Zhang X, Wu~Y.
\newblock Eventanchor: Reducing human interactions in event annotation of racket sports videos.
\newblock In: Proceedings of the 2021 CHI Conference on Human Factors in Computing Systems.
\newblock 2021,  1--13

\bibitem{braylan2021aggregating}
Braylan A, Lease M.
\newblock Aggregating complex annotations via merging and matching.
\newblock In: Proceedings of the 27th ACM SIGKDD Conference on Knowledge Discovery \& Data Mining.
\newblock 2021,  86--94

\bibitem{dumitrache2018capturing}
Dumitrache A, Aroyo L, Welty C.
\newblock Capturing ambiguity in crowdsourcing frame disambiguation.
\newblock In: Sixth AAAI Conference on Human Computation and Crowdsourcing.
\newblock 2018

\bibitem{deng2021multistep}
Deng Z, Xiang Y.
\newblock Multistep planning for crowdsourcing complex consensus tasks.
\newblock Knowledge-Based Systems, 2021, 231: 107447

\bibitem{michael2017crowdsourcing}
Michael J, Stanovsky G, He~L, Dagan I, Zettlemoyer L.
\newblock Crowdsourcing question-answer meaning representations.
\newblock arXiv preprint arXiv:1711.05885, 2017

\bibitem{ipeirotis2014repeated}
Ipeirotis P~G, Provost F, Sheng V~S, Wang J.
\newblock Repeated labeling using multiple noisy labelers.
\newblock Data Mining and Knowledge Discovery, 2014, 28(2): 402--441

\bibitem{abraham2016many}
Abraham I, Alonso O, Kandylas V, Patel R, Shelford S, Slivkins A.
\newblock How many workers to ask? adaptive exploration for collecting high quality labels.
\newblock In: Proceedings of the 39th International ACM SIGIR conference on Research and Development in Information Retrieval.
\newblock 2016,  473--482

\bibitem{lin2014re}
Lin C~H, Weld D~S, others .
\newblock To re (label), or not to re (label).
\newblock In: Second AAAI conference on human computation and crowdsourcing.
\newblock 2014

\bibitem{schmitz2018online}
Schmitz H, Lykourentzou I.
\newblock Online sequencing of non-decomposable macrotasks in expert crowdsourcing.
\newblock ACM Transactions on Social Computing, 2018, 1(1): 1--33

\bibitem{biemann2013creating}
Biemann C.
\newblock Creating a system for lexical substitutions from scratch using crowdsourcing.
\newblock Language Resources and Evaluation, 2013, 47(1): 97--122

\bibitem{bragg2018sprout}
Bragg J, Weld D~S.
\newblock Sprout: Crowd-powered task design for crowdsourcing.
\newblock In: Proceedings of the 31st annual acm symposium on user interface software and technology.
\newblock 2018,  165--176

\bibitem{gadiraju2019crowd}
Gadiraju U, Demartini G, Kawase R, Dietze S.
\newblock Crowd anatomy beyond the good and bad: Behavioral traces for crowd worker modeling and pre-selection.
\newblock Computer Supported Cooperative Work (CSCW), 2019, 28(5): 815--841

\bibitem{tran2015crowdsourcing}
Tran-Thanh L, Huynh T~D, Rosenfeld A, Ramchurn S~D, Jennings N~R.
\newblock Crowdsourcing complex workflows under budget constraints.
\newblock In: Twenty-Ninth AAAI Conference on Artificial Intelligence.
\newblock 2015

\bibitem{boer2016pplib}
Boer P~M~D, Bernstein A.
\newblock Pplib: Toward the automated generation of crowd computing programs using process recombination and auto-experimentation.
\newblock ACM Transactions on Intelligent Systems and Technology (TIST), 2016, 7(4): 1--20

\bibitem{de2017efficiently}
De~Boer P~M, Bernstein A.
\newblock Efficiently identifying a well-performing crowd process for a given problem.
\newblock In: Proceedings of the 2017 ACM Conference on Computer Supported Cooperative Work and Social Computing.
\newblock 2017,  1688--1699

\bibitem{bhuiyan2020investigating}
Bhuiyan M~M, Zhang A~X, Sehat C~M, Mitra T.
\newblock Investigating differences in crowdsourced news credibility assessment: Raters, tasks, and expert criteria.
\newblock Proceedings of the ACM on Human-Computer Interaction, 2020, 4(CSCW2): 1--26

\bibitem{suzuki2018crowdsheet}
Suzuki R, Sakaguchi T, Matsubara M, Kitagawa H, Morishima A.
\newblock Crowdsheet: instant implementation and out-of-hand execution of complex crowdsourcing.
\newblock In: 2018 IEEE 34th International Conference on Data Engineering (ICDE).
\newblock 2018,  1633--1636

\bibitem{dunnmon2020cross}
Dunnmon J~A, Ratner A~J, Saab K, Khandwala N, Markert M, Sagreiya H, Goldman R, Lee-Messer C, Lungren M~P, Rubin D~L, others .
\newblock Cross-modal data programming enables rapid medical machine learning.
\newblock Patterns, 2020, 1(2): 100019

\bibitem{kittur2012crowdweaver}
Kittur A, Khamkar S, Andr{\'e} P, Kraut R.
\newblock Crowdweaver: visually managing complex crowd work.
\newblock In: Proceedings of the ACM 2012 Conference on Computer Supported Cooperative Work.
\newblock 2012,  1033--1036

\bibitem{wang2013quality}
Wang J, Ipeirotis P~G, Provost F.
\newblock Quality-based pricing for crowdsourced workers.
\newblock 2013

\bibitem{cheng2015measuring}
Cheng J, Teevan J, Bernstein M~S.
\newblock Measuring crowdsourcing effort with error-time curves.
\newblock In: Proceedings of the 33rd Annual ACM Conference on Human Factors in Computing Systems.
\newblock 2015,  1365--1374

\bibitem{yin2013effects}
Yin M, Chen Y, Sun Y~A.
\newblock The effects of performance-contingent financial incentives in online labor markets.
\newblock In: Twenty-Seventh AAAI Conference on Artificial Intelligence.
\newblock 2013

\bibitem{cui2017complex}
Cui L, Zhao X, Liu L, Yu~H, Miao Y.
\newblock Complex crowdsourcing task allocation strategies employing supervised and reinforcement learning.
\newblock International Journal of Crowd Science, 2017

\bibitem{he2018task}
He~W, Cui L, Huang C.
\newblock Task assignments in complex collaborative crowdsourcing.
\newblock In: CCF Conference on Computer Supported Cooperative Work and Social Computing.
\newblock 2018,  574--580

\bibitem{kim2016storia}
Kim J, Monroy-Hernandez A.
\newblock Storia: Summarizing social media content based on narrative theory using crowdsourcing.
\newblock In: Proceedings of the 19th ACM Conference on Computer-Supported Cooperative Work \& Social Computing.
\newblock 2016,  1018--1027

\bibitem{nebeling2016wearwrite}
Nebeling M, To~A, Guo A, Freitas d~A~A, Teevan J, Dow S~P, Bigham J~P.
\newblock Wearwrite: Crowd-assisted writing from smartwatches.
\newblock In: Proceedings of the 2016 CHI conference on human factors in computing systems.
\newblock 2016,  3834--3846

\bibitem{hahn2016knowledge}
Hahn N, Chang J, Kim J~E, Kittur A.
\newblock The knowledge accelerator: Big picture thinking in small pieces.
\newblock In: Proceedings of the 2016 CHI Conference on Human Factors in Computing Systems.
\newblock 2016,  2258--2270

\bibitem{mahyar2018communitycrit}
Mahyar N, James M~R, Ng~M~M, Wu~R~A, Dow S~P.
\newblock Communitycrit: inviting the public to improve and evaluate urban design ideas through micro-activities.
\newblock In: Proceedings of the 2018 CHI Conference on Human Factors in Computing Systems.
\newblock 2018,  1--14

\bibitem{wang2018exploring}
Wang N~C, Hicks D, Luther K.
\newblock Exploring trade-offs between learning and productivity in crowdsourced history.
\newblock Proceedings of the ACM on Human-Computer Interaction, 2018, 2(CSCW): 1--24

\bibitem{mcinnis2018crafting}
McInnis B, Leshed G, Cosley D.
\newblock Crafting policy discussion prompts as a task for newcomers.
\newblock Proceedings of the ACM on Human-Computer Interaction, 2018, 2(CSCW): 1--23

\bibitem{allahbakhsh2015task}
Allahbakhsh M, Arbabi S, Shirazi M, Motahari-Nezhad H~R.
\newblock A task decomposition framework for surveying the crowd contextual insights.
\newblock In: 2015 IEEE 8th International Conference on Service-Oriented Computing and Applications (SOCA).
\newblock 2015,  155--162

\bibitem{chilton2013cascade}
Chilton L~B, Little G, Edge D, Weld D~S, Landay J~A.
\newblock Cascade: Crowdsourcing taxonomy creation.
\newblock In: Proceedings of the SIGCHI Conference on Human Factors in Computing Systems.
\newblock 2013,  1999--2008

\bibitem{agapie2015crowdsourcing}
Agapie E, Teevan J, Monroy-Hern{\'a}ndez A.
\newblock Crowdsourcing in the field: A case study using local crowds for event reporting.
\newblock In: Third AAAI Conference on Human Computation and Crowdsourcing.
\newblock 2015

\bibitem{lasecki2013chorus}
Lasecki W~S, Wesley R, Nichols J, Kulkarni A, Allen J~F, Bigham J~P.
\newblock Chorus: a crowd-powered conversational assistant.
\newblock In: Proceedings of the 26th annual ACM symposium on User interface software and technology.
\newblock 2013,  151--162

\bibitem{li2018crowdia}
Li~T, Luther K, North C.
\newblock Crowdia: Solving mysteries with crowdsourced sensemaking.
\newblock Proceedings of the ACM on Human-Computer Interaction, 2018, 2(CSCW): 1--29

\bibitem{zhu2014reviewing}
Zhu H, Dow S~P, Kraut R~E, Kittur A.
\newblock Reviewing versus doing: Learning and performance in crowd assessment.
\newblock In: Proceedings of the 17th ACM conference on Computer supported cooperative work \& social computing.
\newblock 2014,  1445--1455

\bibitem{drapeau2016microtalk}
Drapeau R, Chilton L, Bragg J, Weld D.
\newblock Microtalk: Using argumentation to improve crowdsourcing accuracy.
\newblock In: Proceedings of the AAAI Conference on Human Computation and Crowdsourcing.
\newblock 2016,  32--41

\bibitem{xiong2018smartcrowd}
Xiong T, Yu~Y, Pan M, Yang J.
\newblock Smartcrowd: a workflow framework for complex crowdsourcing tasks.
\newblock In: International Conference on Business Process Management.
\newblock 2018,  387--398

\bibitem{rahman2019optimized}
Rahman H, Roy S~B, Thirumuruganathan S, Amer-Yahia S, Das G.
\newblock Optimized group formation for solving collaborative tasks.
\newblock The VLDB Journal, 2019, 28(1): 1--23

\bibitem{demartini2012zencrowd}
Demartini G, Difallah D~E, Cudr{\'e}-Mauroux P.
\newblock Zencrowd: leveraging probabilistic reasoning and crowdsourcing techniques for large-scale entity linking.
\newblock In: Proceedings of the 21st international conference on World Wide Web.
\newblock 2012,  469--478

\bibitem{aydin2014crowdsourcing}
Aydin B~I, Yilmaz Y~S, Li~Y, Li~Q, Gao J, Demirbas M.
\newblock Crowdsourcing for multiple-choice question answering.
\newblock In: AAAI.
\newblock 2014,  2946--2953

\bibitem{kim2012bayesian}
Kim H~C, Ghahramani Z.
\newblock Bayesian classifier combination.
\newblock In: Artificial Intelligence and Statistics.
\newblock 2012,  619--627

\bibitem{venanzi2014community}
Venanzi M, Guiver J, Kazai G, Kohli P, Shokouhi M.
\newblock Community-based bayesian aggregation models for crowdsourcing.
\newblock In: Proceedings of the 23rd international conference on World wide web.
\newblock 2014,  155--164

\bibitem{heim2017clickstream}
Heim E, Seitel A, Andrulis J, Isensee F, Stock C, Ross T, Maier-Hein L.
\newblock Clickstream analysis for crowd-based object segmentation with confidence.
\newblock IEEE transactions on pattern analysis and machine intelligence, 2017, 40(12): 2814--2826

\bibitem{rechkemmer2021exploring}
Rechkemmer A, Yin M.
\newblock Exploring the effects of goal setting when training for complex crowdsourcing tasks.
\newblock In: IJCAI.
\newblock 2021,  4819--4823

\bibitem{rechkemmer2020motivating}
Rechkemmer A, Yin M.
\newblock Motivating novice crowd workers through goal setting: An investigation into the effects on complex crowdsourcing task training.
\newblock In: Proceedings of the AAAI Conference on Human Computation and Crowdsourcing.
\newblock 2020,  122--131

\bibitem{wang2021teaching}
Wang Z, Sun H.
\newblock Teaching active human learners.
\newblock In: Proceedings of the AAAI Conference on Artificial Intelligence.
\newblock 2021,  5850--5857

\bibitem{zhu2015machine}
Zhu X.
\newblock Machine teaching: An inverse problem to machine learning and an approach toward optimal education.
\newblock In: Proceedings of the AAAI Conference on Artificial Intelligence.
\newblock 2015

\bibitem{abbas2020trainbot}
Abbas T, Khan V~J, Gadiraju U, Markopoulos P.
\newblock Trainbot: A conversational interface to train crowd workers for delivering on-demand therapy.
\newblock In: Proceedings of the AAAI Conference on Human Computation and Crowdsourcing.
\newblock 2020,  3--12

\bibitem{doroudi2016toward}
Doroudi S, Kamar E, Brunskill E, Horvitz E.
\newblock Toward a learning science for complex crowdsourcing tasks.
\newblock In: Proceedings of the 2016 CHI Conference on Human Factors in Computing Systems.
\newblock 2016,  2623--2634

\bibitem{chen2019cicero}
Chen Q, Bragg J, Chilton L~B, Weld D~S.
\newblock Cicero: Multi-turn, contextual argumentation for accurate crowdsourcing.
\newblock In: Proceedings of the 2019 CHI Conference on Human Factors in Computing Systems.
\newblock 2019,  1--14

\bibitem{tang2019leveraging}
Tang W, Yin M, Ho~C~J.
\newblock Leveraging peer communication to enhance crowdsourcing.
\newblock In: The World Wide Web Conference.
\newblock 2019,  1794--1805

\bibitem{liu2016effective}
Liu A, Soderland S, Bragg J, Lin C~H, Ling X, Weld D~S.
\newblock Effective crowd annotation for relation extraction.
\newblock In: Proceedings of the 2016 conference of the North American chapter of the association for computational linguistics: human language technologies.
\newblock 2016,  897--906

\bibitem{verroios2014context}
Verroios V, Bernstein M~S.
\newblock Context trees: Crowdsourcing global understanding from local views.
\newblock In: Second AAAI Conference on Human Computation and Crowdsourcing.
\newblock 2014

\bibitem{willett2012strategies}
Willett W, Heer J, Agrawala M.
\newblock Strategies for crowdsourcing social data analysis.
\newblock In: Proceedings of the SIGCHI Conference on Human Factors in Computing Systems.
\newblock 2012,  227--236

\bibitem{hettiachchi2021challenge}
Hettiachchi D, Schaekermann M, McKinney T~J, Lease M.
\newblock The challenge of variable effort crowdsourcing and how visible gold can help.
\newblock Proceedings of the ACM on Human-Computer Interaction, 2021, 5(CSCW2): 1--26

\bibitem{zhang2017wikum}
Zhang A~X, Verou L, Karger D.
\newblock Wikum: Bridging discussion forums and wikis using recursive summarization.
\newblock In: Proceedings of the 2017 ACM Conference on Computer Supported Cooperative Work and Social Computing.
\newblock 2017,  2082--2096

\bibitem{kaur2018creating}
Kaur H, Williams A~C, Thompson A~L, Lasecki W~S, Iqbal S~T, Teevan J.
\newblock Creating better action plans for writing tasks via vocabulary-based planning.
\newblock Proceedings of the ACM on Human-Computer Interaction, 2018, 2(CSCW): 1--22

\bibitem{wang2021examination}
Wang Y, Papangelis K, Saker M, Lykourentzou I, Khan V~J, Chamberlain A, Grudin J.
\newblock An examination of the work practices of crowdfarms.
\newblock In: Proceedings of the 2021 CHI Conference on Human Factors in Computing Systems.
\newblock 2021,  1--14

\bibitem{huang2020heteroglossia}
Huang C~Y, Huang S~H, Huang T~H~K.
\newblock Heteroglossia: In-situ story ideation with the crowd.
\newblock In: Proceedings of the 2020 CHI Conference on Human Factors in Computing Systems.
\newblock 2020,  1--12

\bibitem{hung2017computing}
Hung N~Q~V, Viet H~H, Tam N~T, Weidlich M, Yin H, Zhou X.
\newblock Computing crowd consensus with partial agreement.
\newblock IEEE Transactions on Knowledge and Data Engineering, 2017, 30(1): 1--14

\bibitem{pei2021quality}
Pei W, Yang Z, Chen M, Yue C.
\newblock Quality control in crowdsourcing based on fine-grained behavioral features.
\newblock Proceedings of the ACM on Human-Computer Interaction, 2021, 5(CSCW2): 1--28

\bibitem{tang2020optimal}
Tang F.
\newblock Optimal complex task assignment in service crowdsourcing.
\newblock In: IJCAI.
\newblock 2020,  1563--1569

\bibitem{mavridis2016using}
Mavridis P, Gross-Amblard D, Mikl{\'o}s Z.
\newblock Using hierarchical skills for optimized task assignment in knowledge-intensive crowdsourcing.
\newblock In: Proceedings of the 25th International Conference on World Wide Web.
\newblock 2016,  843--853

\bibitem{maarry2014skill}
Maarry K~E, Balke W~T, Cho H, Hwang S~w, Baba Y.
\newblock Skill ontology-based model for quality assurance in crowdsourcing.
\newblock In: International conference on database systems for advanced applications.
\newblock 2014,  376--387

\bibitem{hettiachchi2020crowdcog}
Hettiachchi D, Van~Berkel N, Kostakos V, Goncalves J.
\newblock Crowdcog: A cognitive skill based system for heterogeneous task assignment and recommendation in crowdsourcing.
\newblock Proceedings of the ACM on Human-Computer Interaction, 2020, 4(CSCW2): 1--22

\bibitem{aris2019review}
Aris H, Azizan A.
\newblock A review on the methods to evaluate crowd contributions in crowdsourcing applications.
\newblock In: International Conference of Reliable Information and Communication Technology.
\newblock 2019,  1031--1041

\bibitem{zlabinger2020dexa}
Zlabinger M, Sabou M, Hofst{\"a}tter S, Sertkan M, Hanbury A.
\newblock Dexa: Supporting non-expert annotators with dynamic examples from experts.
\newblock In: Proceedings of the 43rd International ACM SIGIR Conference on Research and Development in Information Retrieval.
\newblock 2020,  2109--2112

\bibitem{xu2014voyant}
Xu~A, Huang S~W, Bailey B.
\newblock Voyant: generating structured feedback on visual designs using a crowd of non-experts.
\newblock In: Proceedings of the 17th ACM conference on Computer supported cooperative work \& social computing.
\newblock 2014,  1433--1444

\bibitem{li2020crowdsourced}
Li~J.
\newblock Crowdsourced text sequence aggregation based on hybrid reliability and representation.
\newblock In: Proceedings of the 43rd International ACM SIGIR Conference on Research and Development in Information Retrieval.
\newblock 2020,  1761--1764

\bibitem{braylan2020modeling}
Braylan A, Lease M.
\newblock Modeling and aggregation of complex annotations via annotation distances.
\newblock In: Proceedings of The Web Conference 2020.
\newblock 2020,  1807--1818

\bibitem{de2017crowd}
De~Boer P.
\newblock Crowd process design: how to coordinate crowds to solve complex problems.
\newblock PhD thesis, University of Zurich, 2017

\bibitem{salehi2017communicating}
Salehi N, Teevan J, Iqbal S, Kamar E.
\newblock Communicating context to the crowd for complex writing tasks.
\newblock In: Proceedings of the 2017 ACM Conference on Computer Supported Cooperative Work and Social Computing.
\newblock 2017,  1890--1901

\bibitem{xu2015classroom}
Xu~A, Rao H, Dow S~P, Bailey B~P.
\newblock A classroom study of using crowd feedback in the iterative design process.
\newblock In: Proceedings of the 18th ACM conference on computer supported cooperative work \& social computing.
\newblock 2015,  1637--1648

\bibitem{huang2017supporting}
Huang Y~C, Huang J~C, Wang H~C, Hsu J~Y~j.
\newblock Supporting esl writing by prompting crowdsourced structural feedback.
\newblock In: Fifth AAAI Conference on Human Computation and Crowdsourcing.
\newblock 2017

\bibitem{kobayashi2018empirical}
Kobayashi M, Morita H, Matsubara M, Shimizu N, Morishima A.
\newblock An empirical study on short-and long-term effects of self-correction in crowdsourced microtasks.
\newblock In: Sixth AAAI Conference on Human Computation and Crowdsourcing.
\newblock 2018

\bibitem{jayaram2021human}
Jayaram S, Allaway E.
\newblock Human rationales as attribution priors for explainable stance detection.
\newblock In: Proceedings of the 2021 Conference on Empirical Methods in Natural Language Processing.
\newblock 2021,  5540--5554

\bibitem{maninis2018deep}
Maninis K~K, Caelles S, Pont-Tuset J, Van~Gool L.
\newblock Deep extreme cut: From extreme points to object segmentation.
\newblock In: Proceedings of the IEEE Conference on Computer Vision and Pattern Recognition.
\newblock 2018,  616--625

\bibitem{benenson2019large}
Benenson R, Popov S, Ferrari V.
\newblock Large-scale interactive object segmentation with human annotators.
\newblock In: Proceedings of the IEEE/CVF Conference on Computer Vision and Pattern Recognition.
\newblock 2019,  11700--11709

\bibitem{russakovsky2015best}
Russakovsky O, Li~L~J, Fei-Fei L.
\newblock Best of both worlds: human-machine collaboration for object annotation.
\newblock In: Proceedings of the IEEE conference on computer vision and pattern recognition.
\newblock 2015,  2121--2131

\bibitem{branson2017lean}
Branson S, Van~Horn G, Perona P.
\newblock Lean crowdsourcing: Combining humans and machines in an online system.
\newblock In: Proceedings of the IEEE Conference on Computer Vision and Pattern Recognition.
\newblock 2017,  7474--7483

\bibitem{gouravajhala2018eureca}
Gouravajhala S~R, Yim J, Desingh K, Huang Y, Jenkins O~C, Lasecki W~S.
\newblock Eureca: Enhanced understanding of real environments via crowd assistance.
\newblock In: Sixth AAAI conference on human computation and crowdsourcing.
\newblock 2018

\bibitem{kanchinadam2020rationale}
Kanchinadam T, Westpfahl K, You Q, Fung G.
\newblock Rationale-based human-in-the-loop via supervised attention.
\newblock In: DaSH@ KDD.
\newblock 2020

\bibitem{yang2019scalpel}
Yang J, Smirnova A, Yang D, Demartini G, Lu~Y, Cudr{\'e}-Mauroux P.
\newblock Scalpel-cd: leveraging crowdsourcing and deep probabilistic modeling for debugging noisy training data.
\newblock In: The World Wide Web Conference.
\newblock 2019,  2158--2168

\bibitem{nushi2017human}
Nushi B, Kamar E, Horvitz E, Kossmann D.
\newblock On human intellect and machine failures: Troubleshooting integrative machine learning systems.
\newblock In: Thirty-First AAAI Conference on Artificial Intelligence.
\newblock 2017

\bibitem{das2019joint}
Das A~K, Ashrafi A, Ahmmad M.
\newblock Joint cognition of both human and machine for predicting criminal punishment in judicial system.
\newblock In: 2019 IEEE 4th International Conference on Computer and Communication Systems (ICCCS).
\newblock 2019,  36--40

\bibitem{zheng2017truth}
Zheng Y, Li~G, Li~Y, Shan C, Cheng R.
\newblock Truth inference in crowdsourcing: Is the problem solved?
\newblock Proceedings of the VLDB Endowment, 2017, 10(5): 541--552

\bibitem{li2014resolving}
Li~Q, Li~Y, Gao J, Zhao B, Fan W, Han J.
\newblock Resolving conflicts in heterogeneous data by truth discovery and source reliability estimation.
\newblock In: Proceedings of the 2014 ACM SIGMOD international conference on Management of data.
\newblock 2014,  1187--1198

\bibitem{li2014confidence}
Li~Q, Li~Y, Gao J, Su~L, Zhao B, Demirbas M, Fan W, Han J.
\newblock A confidence-aware approach for truth discovery on long-tail data.
\newblock Proceedings of the VLDB Endowment, 2014, 8(4): 425--436

\bibitem{zhou2012learning}
Zhou D, Basu S, Mao Y, Platt J.
\newblock Learning from the wisdom of crowds by minimax entropy.
\newblock Advances in neural information processing systems, 2012, 25

\bibitem{sucar2015probabilistic}
Sucar L~E.
\newblock Probabilistic graphical models.
\newblock Advances in Computer Vision and Pattern Recognition. London: Springer London. doi, 2015, 10(978): 1

\bibitem{dawid1979maximum}
Dawid A~P, Skene A~M.
\newblock Maximum likelihood estimation of observer error-rates using the em algorithm.
\newblock Journal of the Royal Statistical Society: Series C (Applied Statistics), 1979, 28(1): 20--28

\bibitem{goncalves2017eliciting}
Goncalves J, Hosio S, Kostakos V.
\newblock Eliciting structured knowledge from situated crowd markets.
\newblock ACM Transactions on Internet Technology (TOIT), 2017, 17(2): 1--21

\bibitem{barbosa2019rehumanized}
Barbosa N~M, Chen M.
\newblock Rehumanized crowdsourcing: A labeling framework addressing bias and ethics in machine learning.
\newblock In: Proceedings of the 2019 CHI Conference on Human Factors in Computing Systems.
\newblock 2019,  1--12

\bibitem{simpson2015language}
Simpson E~D, Venanzi M, Reece S, Kohli P, Guiver J, Roberts S~J, Jennings N~R.
\newblock Language understanding in the wild: Combining crowdsourcing and machine learning.
\newblock In: Proceedings of the 24th international conference on world wide web.
\newblock 2015,  992--1002

\bibitem{mutton2007gleu}
Mutton A, Dras M, Wan S, Dale R.
\newblock Gleu: Automatic evaluation of sentence-level fluency.
\newblock In: Proceedings of the 45th Annual Meeting of the Association of Computational Linguistics.
\newblock 2007,  344--351

\bibitem{papineni2002bleu}
Papineni K, Roukos S, Ward T, Zhu W~J.
\newblock Bleu: a method for automatic evaluation of machine translation.
\newblock In: Proceedings of the 40th annual meeting of the Association for Computational Linguistics.
\newblock 2002,  311--318

\bibitem{yu2016unitbox}
Yu~J, Jiang Y, Wang Z, Cao Z, Huang T.
\newblock Unitbox: An advanced object detection network.
\newblock In: Proceedings of the 24th ACM international conference on Multimedia.
\newblock 2016,  516--520

\bibitem{uma2021learning}
Uma A~N, Fornaciari T, Hovy D, Paun S, Plank B, Poesio M.
\newblock Learning from disagreement: A survey.
\newblock Journal of Artificial Intelligence Research, 2021, 72: 1385--1470

\bibitem{baba2020crowdea}
Baba Y, Li~J, Kashima H.
\newblock Crowdea: Multi-view idea prioritization with crowds.
\newblock In: Proceedings of the AAAI Conference on Human Computation and Crowdsourcing.
\newblock 2020,  23--32

\bibitem{vittayakorn2011quality}
Vittayakorn S, Hays J.
\newblock Quality assessment for crowdsourced object annotations.
\newblock In: BMVC.
\newblock 2011,  1--11

\bibitem{chung2019efficient}
Chung J~J~Y, Song J~Y, Kutty S, Hong S, Kim J, Lasecki W~S.
\newblock Efficient elicitation approaches to estimate collective crowd answers.
\newblock Proceedings of the ACM on Human-Computer Interaction, 2019, 3(CSCW): 1--25

\bibitem{li2019dataset}
Li~J, Fukumoto F.
\newblock A dataset of crowdsourced word sequences: Collections and answer aggregation for ground truth creation.
\newblock In: Proceedings of the First Workshop on Aggregating and Analysing Crowdsourced Annotations for NLP.
\newblock 2019,  24--28

\bibitem{li2021label}
Li~J, Endo L~R, Kashima H.
\newblock Label aggregation for crowdsourced triplet similarity comparisons.
\newblock In: International Conference on Neural Information Processing.
\newblock 2021,  176--185

\bibitem{li2022context}
Li~J.
\newblock Context-based collective preference aggregation for prioritizing crowd opinions in social decision-making.
\newblock In: Proceedings of the ACM Web Conference 2022.
\newblock 2022,  2657--2667

\bibitem{timmermans2016exploiting}
Timmermans B.
\newblock Exploiting disagreement through open-ended tasks for capturing interpretation spaces.
\newblock In: European Semantic Web Conference.
\newblock 2016,  873--882

\bibitem{klebanov2014difficult}
Klebanov B~B, Beigman E.
\newblock Difficult cases: From data to learning, and back.
\newblock In: Proceedings of the 52nd Annual Meeting of the Association for Computational Linguistics (Volume 2: Short Papers).
\newblock 2014,  390--396

\bibitem{basile2021we}
Basile V, Fell M, Fornaciari T, Hovy D, Paun S, Plank B, Poesio M, Uma A, others .
\newblock We need to consider disagreement in evaluation.
\newblock In: 1st Workshop on Benchmarking: Past, Present and Future.
\newblock 2021,  15--21

\bibitem{inel2014crowdtruth}
Inel O, Khamkham K, Cristea T, Dumitrache A, Rutjes A, Ploeg J~v~d, Romaszko L, Aroyo L, Sips R~J.
\newblock Crowdtruth: Machine-human computation framework for harnessing disagreement in gathering annotated data.
\newblock In: International semantic web conference.
\newblock 2014,  486--504

\bibitem{sofiiuk2020f}
Sofiiuk K, Petrov I, Barinova O, Konushin A.
\newblock f-brs: Rethinking backpropagating refinement for interactive segmentation.
\newblock In: Proceedings of the IEEE/CVF Conference on Computer Vision and Pattern Recognition.
\newblock 2020,  8623--8632

\bibitem{lan2019learning}
Lan O, Huang X, Lin B~Y, Jiang H, Liu L, Ren X.
\newblock Learning to contextually aggregate multi-source supervision for sequence labeling.
\newblock arXiv preprint arXiv:1910.04289, 2019

\bibitem{sameki2015characterizing}
Sameki M, Gurari D, Betke M.
\newblock Characterizing image segmentation behavior of the crowd.
\newblock Collective Intelligence, 2015,  1--4

\bibitem{jang2019interactive}
Jang W~D, Kim C~S.
\newblock Interactive image segmentation via backpropagating refinement scheme.
\newblock In: Proceedings of the IEEE/CVF Conference on Computer Vision and Pattern Recognition.
\newblock 2019,  5297--5306

\bibitem{yang2022incorporating}
Yang Y, Chen P, Sun H.
\newblock Incorporating pixel proximity into answer aggregation for crowdsourced image segmentation.
\newblock CCF Transactions on Pervasive Computing and Interaction, 2022,  1--16

\bibitem{flower1981cognitive}
Flower L, Hayes J~R.
\newblock A cognitive process theory of writing.
\newblock College composition and communication, 1981, 32(4): 365--387

\bibitem{russell1993cost}
Russell D~M, Stefik M~J, Pirolli P, Card S~K.
\newblock The cost structure of sensemaking.
\newblock In: Proceedings of the INTERACT'93 and CHI'93 conference on Human factors in computing systems.
\newblock 1993,  269--276

\bibitem{bradel2013large}
Bradel L, Endert A, Koch K, Andrews C, North C.
\newblock Large high resolution displays for co-located collaborative sensemaking: Display usage and territoriality.
\newblock International Journal of Human-Computer Studies, 2013, 71(11): 1078--1088

\bibitem{lasecki2015apparition}
Lasecki W~S, Kim J, Rafter N, Sen O, Bigham J~P, Bernstein M~S.
\newblock Apparition: Crowdsourced user interfaces that come to life as you sketch them.
\newblock In: Proceedings of the 33rd Annual ACM Conference on Human Factors in Computing Systems.
\newblock 2015,  1925--1934

\bibitem{rappaz2018latent}
Rappaz J, Catasta M, West R, Aberer K.
\newblock Latent structure in collaboration: the case of reddit r/place.
\newblock In: Twelfth International AAAI Conference on Web and Social Media.
\newblock 2018

\bibitem{feldman2021we}
Feldman M~Q, McInnis B~J.
\newblock How we write with crowds.
\newblock Proceedings of the ACM on Human-Computer Interaction, 2021, 4(CSCW3): 1--31

\bibitem{kraut1992task}
Kraut R, Galegher J, Fish R, Chalfonte B.
\newblock Task requirements and media choice in collaborative writing.
\newblock Human--Computer Interaction, 1992, 7(4): 375--407

\bibitem{glassman2016learnersourcing}
Glassman E~L, Lin A, Cai C~J, Miller R~C.
\newblock Learnersourcing personalized hints.
\newblock In: Proceedings of the 19th ACM conference on computer-supported cooperative work \& social computing.
\newblock 2016,  1626--1636

\bibitem{kim2017mechanical}
Kim J, Sterman S, Cohen A~A~B, Bernstein M~S.
\newblock Mechanical novel: Crowdsourcing complex work through reflection and revision.
\newblock In: Proceedings of the 2017 acm conference on computer supported cooperative work and social computing.
\newblock 2017,  233--245

\bibitem{luther2015structuring}
Luther K, Tolentino J~L, Wu~W, Pavel A, Bailey B~P, Agrawala M, Hartmann B, Dow S~P.
\newblock Structuring, aggregating, and evaluating crowdsourced design critique.
\newblock In: Proceedings of the 18th ACM conference on computer supported cooperative work \& social computing.
\newblock 2015,  473--485

\bibitem{lin2014crowdsourced}
Lin C~C, Huang Y~C, Hsu J~Y~j.
\newblock Crowdsourced explanations for humorous internet memes based on linguistic theories.
\newblock In: Second AAAI Conference on Human Computation and Crowdsourcing.
\newblock 2014

\bibitem{pirolli2005sensemaking}
Pirolli P, Card S.
\newblock The sensemaking process and leverage points for analyst technology as identified through cognitive task analysis.
\newblock In: Proceedings of international conference on intelligence analysis.
\newblock 2005,  2--4

\bibitem{ericsson1996expert}
Ericsson K~A, Lehmann A~C.
\newblock Expert and exceptional performance: Evidence of maximal adaptation to task constraints.
\newblock Annual review of psychology, 1996, 47(1): 273--305

\bibitem{cheng2009context}
Cheng W~H, Gotz D.
\newblock Context-based page unit recommendation for web-based sensemaking tasks.
\newblock In: Proceedings of the 14th international conference on Intelligent user interfaces.
\newblock 2009,  107--116

\bibitem{drieger2013semantic}
Drieger P.
\newblock Semantic network analysis as a method for visual text analytics.
\newblock Procedia-social and behavioral sciences, 2013, 79: 4--17

\bibitem{sun2015biset}
Sun M, Mi~P, North C, Ramakrishnan N.
\newblock Biset: Semantic edge bundling with biclusters for sensemaking.
\newblock IEEE transactions on visualization and computer graphics, 2015, 22(1): 310--319

\bibitem{zhao2017supporting}
Zhao J, Glueck M, Isenberg P, Chevalier F, Khan A.
\newblock Supporting handoff in asynchronous collaborative sensemaking using knowledge-transfer graphs.
\newblock IEEE transactions on visualization and computer graphics, 2017, 24(1): 340--350

\bibitem{tavanapour2017collaboration}
Tavanapour N, Bittner E~A~C.
\newblock Collaboration among crowdsourcees: Towards a design theory for collaboration process design.
\newblock 2017

\bibitem{madge2019crowdsourcing}
Madge C, Yu~J, Chamberlain J, Kruschwitz U, Paun S, Poesio M.
\newblock Crowdsourcing and aggregating nested markable annotations.
\newblock 2019

\bibitem{guillaume2016crowdsourcing}
Guillaume B, Fort K, Lefebvre N.
\newblock Crowdsourcing complex language resources: Playing to annotate dependency syntax.
\newblock In: International Conference on Computational Linguistics (COLING).
\newblock 2016

\bibitem{urbanek2019learning}
Urbanek J, Fan A, Karamcheti S, Jain S, Humeau S, Dinan E, Rockt{\"a}schel T, Kiela D, Szlam A, Weston J.
\newblock Learning to speak and act in a fantasy text adventure game.
\newblock arXiv preprint arXiv:1903.03094, 2019

\bibitem{muldoon2018survey}
Muldoon C, O’Grady M~J, O’Hare G~M.
\newblock A survey of incentive engineering for crowdsourcing.
\newblock The Knowledge Engineering Review, 2018, 33

\bibitem{scekic2015supporting}
Scekic O, Truong H~L, Dustdar S.
\newblock Supporting multilevel incentive mechanisms in crowdsourcing systems: an artifact-centric view.
\newblock In: Crowdsourcing,  91--111. Springer, 2015

\bibitem{lee2018aggregating}
Lee D, Das~Sarma A, Parameswaran A.
\newblock Aggregating crowdsourced image segmentations.
\newblock HCOMP, 2018

\bibitem{braylan2022measuring}
Braylan A, Alonso O, Lease M.
\newblock Measuring annotator agreement generally across complex structured, multi-object, and free-text annotation tasks.
\newblock In: Proceedings of the ACM Web Conference 2022.
\newblock 2022,  1720--1730

\bibitem{gadiraju2020can}
Gadiraju U, Yang J.
\newblock What can crowd computing do for the next generation of ai systems?
\newblock In: CSW@ NeurIPS.
\newblock 2020,  7--13

\bibitem{lu2007survey}
Lu~D, Weng Q.
\newblock A survey of image classification methods and techniques for improving classification performance.
\newblock International journal of Remote sensing, 2007, 28(5): 823--870

\bibitem{medhat2014sentiment}
Medhat W, Hassan A, Korashy H.
\newblock Sentiment analysis algorithms and applications: A survey.
\newblock Ain Shams engineering journal, 2014, 5(4): 1093--1113

\bibitem{mithe2013optical}
Mithe R, Indalkar S, Divekar N.
\newblock Optical character recognition.
\newblock International journal of recent technology and engineering (IJRTE), 2013, 2(1): 72--75

\bibitem{deng2009imagenet}
Deng J, Dong W, Socher R, Li~L~J, Li~K, Fei-Fei L.
\newblock Imagenet: A large-scale hierarchical image database.
\newblock In: 2009 IEEE conference on computer vision and pattern recognition.
\newblock 2009,  248--255

\bibitem{raykar2010learning}
Raykar V~C, Yu~S, Zhao L~H, Valadez G~H, Florin C, Bogoni L, Moy L.
\newblock Learning from crowds.
\newblock Journal of machine learning research, 2010, 11(4)

\bibitem{krishna2017visual}
Krishna R, Zhu Y, Groth O, Johnson J, Hata K, Kravitz J, Chen S, Kalantidis Y, Li~L~J, Shamma D~A, others .
\newblock Visual genome: Connecting language and vision using crowdsourced dense image annotations.
\newblock International journal of computer vision, 2017, 123(1): 32--73

\bibitem{nanni2020data}
Nanni L, Maguolo G, Paci M.
\newblock Data augmentation approaches for improving animal audio classification.
\newblock Ecological Informatics, 2020, 57: 101084

\bibitem{wexler2001and}
Wexler M~N.
\newblock The who, what and why of knowledge mapping.
\newblock Journal of knowledge management, 2001

\bibitem{shin2015incremental}
Shin J, Wu~S, Wang F, De~Sa C, Zhang C, R{\'e}~C.
\newblock Incremental knowledge base construction using deepdive.
\newblock In: Proceedings of the VLDB Endowment International Conference on Very Large Data Bases.
\newblock 2015,  1310

\bibitem{zhao2003face}
Zhao W, Chellappa R, Phillips P~J, Rosenfeld A.
\newblock Face recognition: A literature survey.
\newblock ACM computing surveys (CSUR), 2003, 35(4): 399--458

\bibitem{chen2019antprophet}
Chen C, Zhang X, Ju~S, Fu~C, Tang C, Zhou J, Li~X.
\newblock Antprophet: an intention mining system behind alipay's intelligent customer service bot.
\newblock In: IJCAI.
\newblock 2019,  6497--6499

\bibitem{gray2008viewpoint}
Gray D, Tao H.
\newblock Viewpoint invariant pedestrian recognition with an ensemble of localized features.
\newblock In: European conference on computer vision.
\newblock 2008,  262--275

\bibitem{chai2022error}
Chai L, Sun H, Wang Z.
\newblock An error consistency based approach to answer aggregation in open-ended crowdsourcing.
\newblock Information Sciences, 2022, 608: 1029--1044

\bibitem{cheng2001color}
Cheng H~D, Jiang X~H, Sun Y, Wang J.
\newblock Color image segmentation: advances and prospects.
\newblock Pattern recognition, 2001, 34(12): 2259--2281

\bibitem{schmidt2016using}
Schmidt G~B, Jettinghoff W~M.
\newblock Using amazon mechanical turk and other compensated crowdsourcing sites.
\newblock Business Horizons, 2016, 59(4): 391--400

\bibitem{levinson2011towards}
Levinson J, Askeland J, Becker J, Dolson J, Held D, Kammel S, Kolter J~Z, Langer D, Pink O, Pratt V, others .
\newblock Towards fully autonomous driving: Systems and algorithms.
\newblock In: 2011 IEEE intelligent vehicles symposium (IV).
\newblock 2011,  163--168

\bibitem{yaqoob2017enabling}
Yaqoob I, Hashem I~A~T, Mehmood Y, Gani A, Mokhtar S, Guizani S.
\newblock Enabling communication technologies for smart cities.
\newblock IEEE Communications Magazine, 2017, 55(1): 112--120

\bibitem{wu2021task}
Wu~G, Chen Z, Liu J, Han D, Qiao B.
\newblock Task assignment for social-oriented crowdsourcing.
\newblock Frontiers of Computer Science, 2021, 15: 1--11

\bibitem{han2021find}
Han T, Sun H, Song Y, Fang Y, Liu X.
\newblock Find truth in the hands of the few: acquiring specific knowledge with crowdsourcing.
\newblock Frontiers of Computer Science, 2021, 15: 1--12

\bibitem{hu2020quality}
Hu~Z, Wu~W, Luo J, Wang X, Li~B.
\newblock Quality assessment in competition-based software crowdsourcing.
\newblock Frontiers of Computer Science, 2020, 14: 1--14

\bibitem{ren2024label}
Ren L, Jiang L, Zhang W, Li~C.
\newblock Label distribution similarity-based noise correction for crowdsourcing.
\newblock Frontiers of Computer Science, 2024, 18(5): 185323

\bibitem{zhang2023attribute}
Zhang Y, Jiang L, Li~C.
\newblock Attribute augmentation-based label integration for crowdsourcing.
\newblock Frontiers of Computer Science, 2023, 17(5): 175331

\bibitem{lee2021all}
Lee L~H, Braud T, Zhou P, Wang L, Xu~D, Lin Z, Kumar A, Bermejo C, Hui P.
\newblock All one needs to know about metaverse: A complete survey on technological singularity, virtual ecosystem, and research agenda.
\newblock arXiv preprint arXiv:2110.05352, 2021

\bibitem{paun2019proceedings}
Paun S, Hovy D.
\newblock Proceedings of the first workshop on aggregating and analysing crowdsourced annotations for nlp.
\newblock In: Proceedings of the First Workshop on Aggregating and Analysing Crowdsourced Annotations for NLP.
\newblock 2019

\bibitem{cer2018universal}
Cer D, Yang Y, Kong S~y, Hua N, Limtiaco N, John R~S, Constant N, Guajardo-Cespedes M, Yuan S, Tar C, others .
\newblock Universal sentence encoder.
\newblock arXiv preprint arXiv:1803.11175, 2018

\bibitem{devlin2018bert}
Devlin J, Chang M~W, Lee K, Toutanova K.
\newblock Bert: Pre-training of deep bidirectional transformers for language understanding.
\newblock arXiv preprint arXiv:1810.04805, 2018

\bibitem{ustalov2021vldb}
Ustalov D, Pavlichenko N, Stelmakh I, Kuznetsov D.
\newblock Vldb 2021 crowd science challenge on aggregating crowdsourced audio transcriptions.
\newblock In: Proceedings of the 2nd Crowd Science Workshop: Trust, Ethics, and Excellence in Crowdsourced Data Management at Scale.
\newblock 2021,  1--7

\bibitem{koller2009probabilistic}
Koller D, Friedman N.
\newblock Probabilistic graphical models: principles and techniques.
\newblock MIT press, 2009

\bibitem{jumper2021highly}
Jumper J, Evans R, Pritzel A, Green T, Figurnov M, Ronneberger O, Tunyasuvunakool K, Bates R, {\v{Z}}{\'\i}dek A, Potapenko A, others .
\newblock Highly accurate protein structure prediction with alphafold.
\newblock Nature, 2021, 596(7873): 583--589

\bibitem{difallah2015dynamics}
Difallah D~E, Catasta M, Demartini G, Ipeirotis P~G, Cudr{\'e}-Mauroux P.
\newblock The dynamics of micro-task crowdsourcing: The case of amazon mturk.
\newblock In: Proceedings of the 24th international conference on world wide web.
\newblock 2015,  238--247

\bibitem{binns2018s}
Binns R, Van~Kleek M, Veale M, Lyngs U, Zhao J, Shadbolt N.
\newblock 'it's reducing a human being to a percentage' perceptions of justice in algorithmic decisions.
\newblock In: Proceedings of the 2018 Chi conference on human factors in computing systems.
\newblock 2018,  1--14

\bibitem{kittur2013future}
Kittur A, Nickerson J~V, Bernstein M, Gerber E, Shaw A, Zimmerman J, Lease M, Horton J.
\newblock The future of crowd work.
\newblock In: Proceedings of the 2013 conference on Computer supported cooperative work.
\newblock 2013,  1301--1318

\bibitem{retelny2017no}
Retelny D, Bernstein M~S, Valentine M~A.
\newblock No workflow can ever be enough: How crowdsourcing workflows constrain complex work.
\newblock Proceedings of the ACM on Human-Computer Interaction, 2017, 1(CSCW): 1--23

\bibitem{howe2006rise}
Howe J, others .
\newblock The rise of crowdsourcing.
\newblock Wired magazine, 2006, 14(6): 1--4

\bibitem{kerner2016failure}
Kerner B~S.
\newblock Failure of classical traffic flow theories: Stochastic highway capacity and automatic driving.
\newblock Physica A: Statistical Mechanics and its Applications, 2016, 450: 700--747

\bibitem{minaee2021image}
Minaee S, Boykov Y~Y, Porikli F, Plaza A~J, Kehtarnavaz N, Terzopoulos D.
\newblock Image segmentation using deep learning: A survey.
\newblock IEEE transactions on pattern analysis and machine intelligence, 2021

\bibitem{vedantam2015cider}
Vedantam R, Lawrence~Zitnick C, Parikh D.
\newblock Cider: Consensus-based image description evaluation.
\newblock In: Proceedings of the IEEE conference on computer vision and pattern recognition.
\newblock 2015,  4566--4575

\bibitem{mesbah2023hybrideval}
Mesbah S, Arous I, Yang J, Bozzon A.
\newblock Hybrideval: A human-ai collaborative approach for evaluating design ideas at scale.
\newblock In: Proceedings of the ACM Web Conference 2023.
\newblock 2023,  3837--3848

\bibitem{arous2020opencrowd}
Arous I, Yang J, Khayati M, Cudr{\'e}-Mauroux P.
\newblock Opencrowd: A human-ai collaborative approach for finding social influencers via open-ended answers aggregation.
\newblock In: Proceedings of The Web Conference 2020.
\newblock 2020,  1851--1862

\bibitem{von2008recaptcha}
Von~Ahn L, Maurer B, McMillen C, Abraham D, Blum M.
\newblock recaptcha: Human-based character recognition via web security measures.
\newblock Science, 2008, 321(5895): 1465--1468

\bibitem{helouetdata}
H{\'e}lou{\"e}t L, Singh R, Miklos Z.
\newblock Data centric workflows for complex crowdsourcing applications

\bibitem{tong2017spatial}
Tong Y, Chen L, Shahabi C.
\newblock Spatial crowdsourcing: Challenges, techniques, and applications.
\newblock Proceedings of the VLDB Endowment, 2017, 10(12): 1988--1991

\bibitem{krizhevsky2017imagenet}
Krizhevsky A, Sutskever I, Hinton G~E.
\newblock Imagenet classification with deep convolutional neural networks.
\newblock Communications of the ACM, 2017, 60(6): 84--90

\bibitem{he2015delving}
He~K, Zhang X, Ren S, Sun J.
\newblock Delving deep into rectifiers: Surpassing human-level performance on imagenet classification.
\newblock In: Proceedings of the IEEE international conference on computer vision.
\newblock 2015,  1026--1034

\bibitem{russakovsky2015imagenet}
Russakovsky O, Deng J, Su~H, Krause J, Satheesh S, Ma~S, Huang Z, Karpathy A, Khosla A, Bernstein M, others .
\newblock Imagenet large scale visual recognition challenge.
\newblock International journal of computer vision, 2015, 115: 211--252

\bibitem{he2016deep}
He~K, Zhang X, Ren S, Sun J.
\newblock Deep residual learning for image recognition.
\newblock In: Proceedings of the IEEE conference on computer vision and pattern recognition.
\newblock 2016,  770--778

\bibitem{zeiler2014visualizing}
Zeiler M~D, Fergus R.
\newblock Visualizing and understanding convolutional networks.
\newblock In: Computer Vision--ECCV 2014: 13th European Conference, Zurich, Switzerland, September 6-12, 2014, Proceedings, Part I 13.
\newblock 2014,  818--833

\bibitem{redmon2017yolo9000}
Redmon J, Farhadi A.
\newblock Yolo9000: better, faster, stronger.
\newblock In: Proceedings of the IEEE conference on computer vision and pattern recognition.
\newblock 2017,  7263--7271

\bibitem{xie2017aggregated}
Xie S, Girshick R, Doll{\'a}r P, Tu~Z, He~K.
\newblock Aggregated residual transformations for deep neural networks.
\newblock In: Proceedings of the IEEE conference on computer vision and pattern recognition.
\newblock 2017,  1492--1500

\bibitem{chollet2017xception}
Chollet F.
\newblock Xception: Deep learning with depthwise separable convolutions.
\newblock In: Proceedings of the IEEE conference on computer vision and pattern recognition.
\newblock 2017,  1251--1258

\bibitem{wang2019latte}
Wang B, Wu~V, Wu~B, Keutzer K.
\newblock Latte: accelerating lidar point cloud annotation via sensor fusion, one-click annotation, and tracking.
\newblock In: 2019 IEEE Intelligent Transportation Systems Conference (ITSC).
\newblock 2019,  265--272

\bibitem{song2019popup}
Song J~Y, Lemmer S~J, Liu M~X, Yan S, Kim J, Corso J~J, Lasecki W~S.
\newblock Popup: reconstructing 3d video using particle filtering to aggregate crowd responses.
\newblock In: Proceedings of the 24th International Conference on Intelligent User Interfaces.
\newblock 2019,  558--569

\bibitem{liu2018joint}
Liu Z, Wang L, Hua G, Zhang Q, Niu Z, Wu~Y, Zheng N.
\newblock Joint video object discovery and segmentation by coupled dynamic markov networks.
\newblock IEEE Transactions on Image Processing, 2018, 27(12): 5840--5853

\bibitem{wang2018segment}
Wang L, Duan X, Zhang Q, Niu Z, Hua G, Zheng N.
\newblock Segment-tube: Spatio-temporal action localization in untrimmed videos with per-frame segmentation.
\newblock Sensors, 2018, 18(5): 1657

\bibitem{chen2022adversarial}
Chen P, Sun H, Yang Y, Chen Z.
\newblock Adversarial learning from crowds.
\newblock In: Proceedings of the AAAI Conference on Artificial Intelligence.
\newblock 2022,  5304--5312

\bibitem{gonzalez2015review}
Gonz{\'a}lez D, P{\'e}rez J, Milan{\'e}s V, Nashashibi F.
\newblock A review of motion planning techniques for automated vehicles.
\newblock IEEE Transactions on intelligent transportation systems, 2015, 17(4): 1135--1145

\bibitem{fortson2012galaxy}
Fortson L, Masters K, Nichol R, Edmondson E, Lintott C, Raddick J, Wallin J.
\newblock Galaxy zoo.
\newblock Advances in machine learning and data mining for astronomy, 2012, 2012: 213--236

\bibitem{khare2016crowdsourcing}
Khare R, Good B~M, Leaman R, Su~A~I, Lu~Z.
\newblock Crowdsourcing in biomedicine: challenges and opportunities.
\newblock Briefings in bioinformatics, 2016, 17(1): 23--32

\bibitem{han2021exploring}
Han J, Brown C, Chauhan J, Grammenos A, Hasthanasombat A, Spathis D, Xia T, Cicuta P, Mascolo C.
\newblock Exploring automatic covid-19 diagnosis via voice and symptoms from crowdsourced data.
\newblock In: ICASSP 2021-2021 IEEE International Conference on Acoustics, Speech and Signal Processing (ICASSP).
\newblock 2021,  8328--8332

\bibitem{cooper2010predicting}
Cooper S, Khatib F, Treuille A, Barbero J, Lee J, Beenen M, Leaver-Fay A, Baker D, Popovi{\'c} Z, players F.
\newblock Predicting protein structures with a multiplayer online game.
\newblock Nature, 2010, 466(7307): 756--760

\bibitem{nuessle2020planning}
Nuessle T~M, McNamara P~A, Garneau N~L.
\newblock Planning and executing scientifically sound community science in a public-facing institution.
\newblock Citizen Science: Theory and Practice, 2020, 5(1)

\bibitem{vaughan2017making}
Vaughan J~W.
\newblock Making better use of the crowd: How crowdsourcing can advance machine learning research.
\newblock J. Mach. Learn. Res., 2017, 18(1): 7026--7071

\bibitem{dellermann2021future}
Dellermann D, Calma A, Lipusch N, Weber T, Weigel S, Ebel P.
\newblock The future of human-ai collaboration: a taxonomy of design knowledge for hybrid intelligence systems.
\newblock arXiv preprint arXiv:2105.03354, 2021

\bibitem{orting2019survey}
{\O}rting S, Doyle A, Hilten v~A, Hirth M, Inel O, Madan C~R, Mavridis P, Spiers H, Cheplygina V.
\newblock A survey of crowdsourcing in medical image analysis.
\newblock arXiv preprint arXiv:1902.09159, 2019

\bibitem{kovashka2016crowdsourcing}
Kovashka A, Russakovsky O, Fei-Fei L, Grauman K, others .
\newblock Crowdsourcing in computer vision.
\newblock Foundations and Trends{\textregistered} in computer graphics and Vision, 2016, 10(3): 177--243

\bibitem{daniel2018quality}
Daniel F, Kucherbaev P, Cappiello C, Benatallah B, Allahbakhsh M.
\newblock Quality control in crowdsourcing: A survey of quality attributes, assessment techniques, and assurance actions.
\newblock ACM Computing Surveys (CSUR), 2018, 51(1): 1--40

\bibitem{tsvetkova2017understanding}
Tsvetkova M, Yasseri T, Meyer E~T, Pickering J~B, Engen V, Walland P, L{\"u}ders M, F{\o}lstad A, Bravos G.
\newblock Understanding human-machine networks: a cross-disciplinary survey.
\newblock ACM Computing Surveys (CSUR), 2017, 50(1): 1--35

\bibitem{jin2020technical}
Jin Y, Carman M, Zhu Y, Xiang Y.
\newblock A technical survey on statistical modelling and design methods for crowdsourcing quality control.
\newblock Artificial Intelligence, 2020, 287: 103351

\bibitem{crosby1979quality}
Crosby P~B.
\newblock Quality is free: The art of making quality certain.
\newblock (No Title), 1979

\bibitem{nguyen2017argument}
Nguyen Q~V~H, Duong C~T, Nguyen T~T, Weidlich M, Aberer K, Yin H, Zhou X.
\newblock Argument discovery via crowdsourcing.
\newblock The VLDB Journal, 2017, 26: 511--535

\bibitem{guo2015mobile}
Guo B, Wang Z, Yu~Z, Wang Y, Yen N~Y, Huang R, Zhou X.
\newblock Mobile crowd sensing and computing: The review of an emerging human-powered sensing paradigm.
\newblock ACM computing surveys (CSUR), 2015, 48(1): 1--31

\bibitem{de2015reliable}
De~Alfaro L, Polychronopoulos V, Shavlovsky M.
\newblock Reliable aggregation of boolean crowdsourced tasks.
\newblock In: Proceedings of the AAAI Conference on Human Computation and Crowdsourcing.
\newblock 2015,  42--51

\bibitem{pavlichenko2021crowdspeech}
Pavlichenko N, Stelmakh I, Ustalov D.
\newblock Crowdspeech and voxdiy: Benchmark datasets for crowdsourced audio transcription.
\newblock arXiv preprint arXiv:2107.01091, 2021

\bibitem{lipping2019crowdsourcing}
Lipping S, Drossos K, Virtanen T.
\newblock Crowdsourcing a dataset of audio captions.
\newblock arXiv preprint arXiv:1907.09238, 2019

\bibitem{kaspar2018crowd}
Kaspar A, Patterson G, Kim C, Aksoy Y, Matusik W, Elgharib M.
\newblock Crowd-guided ensembles: How can we choreograph crowd workers for video segmentation?
\newblock In: Proceedings of the 2018 CHI Conference on Human Factors in Computing Systems.
\newblock 2018,  1--12

\bibitem{singer2013pricing}
Singer Y, Mittal M.
\newblock Pricing mechanisms for crowdsourcing markets.
\newblock In: Proceedings of the 22nd international conference on World Wide Web.
\newblock 2013,  1157--1166

\bibitem{cychnerski2021segmentation}
Cychnerski J, Dziubich T.
\newblock Segmentation quality refinement in large-scale medical image dataset with crowd-sourced annotations.
\newblock In: European Conference on Advances in Databases and Information Systems.
\newblock 2021,  205--216

\bibitem{mo2013cross}
Mo~K, Zhong E, Yang Q.
\newblock Cross-task crowdsourcing.
\newblock In: Proceedings of the 19th ACM SIGKDD international conference on knowledge discovery and data mining.
\newblock 2013,  677--685

\bibitem{khan2017crowddqs}
Khan A~R, Garcia-Molina H.
\newblock Crowddqs: Dynamic question selection in crowdsourcing systems.
\newblock In: Proceedings of the 2017 ACM International Conference on Management of Data.
\newblock 2017,  1447--1462

\bibitem{openai2023gpt}
OpenAI R.
\newblock Gpt-4 technical report.
\newblock arXiv, 2023,  2303--08774

\bibitem{wu2023llms}
Wu~T, Zhu H, Albayrak M, Axon A, Bertsch A, Deng W, Ding Z, Guo B, Gururaja S, Kuo T~S, others .
\newblock Llms as workers in human-computational algorithms? replicating crowdsourcing pipelines with llms.
\newblock arXiv preprint arXiv:2307.10168, 2023

\bibitem{parameswaran2023revisiting}
Parameswaran A~G, Shankar S, Asawa P, Jain N, Wang Y.
\newblock Revisiting prompt engineering via declarative crowdsourcing.
\newblock arXiv preprint arXiv:2308.03854, 2023

\bibitem{he2023annollm}
He~X, Lin Z, Gong Y, Jin A, Zhang H, Lin C, Jiao J, Yiu S~M, Duan N, Chen W, others .
\newblock Annollm: Making large language models to be better crowdsourced annotators.
\newblock arXiv preprint arXiv:2303.16854, 2023

\bibitem{ouyang2022training}
Ouyang L, Wu~J, Jiang X, Almeida D, Wainwright C, Mishkin P, Zhang C, Agarwal S, Slama K, Ray A, others .
\newblock Training language models to follow instructions with human feedback.
\newblock Advances in Neural Information Processing Systems, 2022, 35: 27730--27744

\bibitem{christiano2017deep}
Christiano P~F, Leike J, Brown T, Martic M, Legg S, Amodei D.
\newblock Deep reinforcement learning from human preferences.
\newblock Advances in neural information processing systems, 2017, 30

\bibitem{bai2022training}
Bai Y, Jones A, Ndousse K, Askell A, Chen A, DasSarma N, Drain D, Fort S, Ganguli D, Henighan T, others .
\newblock Training a helpful and harmless assistant with reinforcement learning from human feedback.
\newblock arXiv preprint arXiv:2204.05862, 2022

\bibitem{anil2023palm}
Anil R, Dai A~M, Firat O, Johnson M, Lepikhin D, Passos A, Shakeri S, Taropa E, Bailey P, Chen Z, others .
\newblock Palm 2 technical report.
\newblock arXiv preprint arXiv:2305.10403, 2023

\bibitem{touvron2023llama}
Touvron H, Martin L, Stone K, Albert P, Almahairi A, Babaei Y, Bashlykov N, Batra S, Bhargava P, Bhosale S, others .
\newblock Llama 2: Open foundation and fine-tuned chat models.
\newblock arXiv preprint arXiv:2307.09288, 2023

\bibitem{wu2022ai}
Wu~T, Terry M, Cai C~J.
\newblock Ai chains: Transparent and controllable human-ai interaction by chaining large language model prompts.
\newblock In: Proceedings of the 2022 CHI conference on human factors in computing systems.
\newblock 2022,  1--22

\bibitem{kittur2011crowdforge}
Kittur A, Smus B, Khamkar S, Kraut R~E.
\newblock Crowdforge: Crowdsourcing complex work.
\newblock In: Proceedings of the 24th annual ACM symposium on User interface software and technology.
\newblock 2011,  43--52

\bibitem{bernstein2010soylent}
Bernstein M~S, Little G, Miller R~C, Hartmann B, Ackerman M~S, Karger D~R, Crowell D, Panovich K.
\newblock Soylent: a word processor with a crowd inside.
\newblock In: Proceedings of the 23nd annual ACM symposium on User interface software and technology.
\newblock 2010,  313--322

\bibitem{wang2022self}
Wang X, Wei J, Schuurmans D, Le~Q, Chi E, Narang S, Chowdhery A, Zhou D.
\newblock Self-consistency improves chain of thought reasoning in language models.
\newblock arXiv preprint arXiv:2203.11171, 2022

\bibitem{press2022measuring}
Press O, Zhang M, Min S, Schmidt L, Smith N~A, Lewis M.
\newblock Measuring and narrowing the compositionality gap in language models.
\newblock arXiv preprint arXiv:2210.03350, 2022

\bibitem{shinn2023reflexion}
Shinn N, Cassano F, Labash B, Gopinath A, Narasimhan K, Yao S.
\newblock Reflexion: Language agents with verbal reinforcement learning.
\newblock arXiv preprint arXiv:2303.11366, 2023

\bibitem{veselovsky2023artificial}
Veselovsky V, Ribeiro M~H, West R.
\newblock Artificial artificial artificial intelligence: Crowd workers widely use large language models for text production tasks.
\newblock arXiv preprint arXiv:2306.07899, 2023

\bibitem{marcus2015crowdsourced}
Marcus A, Parameswaran A, others .
\newblock Crowdsourced data management: Industry and academic perspectives.
\newblock Foundations and Trends{\textregistered} in Databases, 2015, 6(1-2): 1--161

\bibitem{rzadca2020autopilot}
Rzadca K, Findeisen P, Swiderski J, Zych P, Broniek P, Kusmierek J, Nowak P, Strack B, Witusowski P, Hand S, others .
\newblock Autopilot: workload autoscaling at google.
\newblock In: Proceedings of the Fifteenth European Conference on Computer Systems.
\newblock 2020,  1--16

\bibitem{hettiachchi2022survey}
Hettiachchi D, Kostakos V, Goncalves J.
\newblock A survey on task assignment in crowdsourcing.
\newblock ACM Computing Surveys (CSUR), 2022, 55(3): 1--35

\end{thebibliography}

\vspace{30pt}
\begin{wrapfigure}{l}{25mm}
\includegraphics[width=1in,height=1.25in,clip,keepaspectratio]{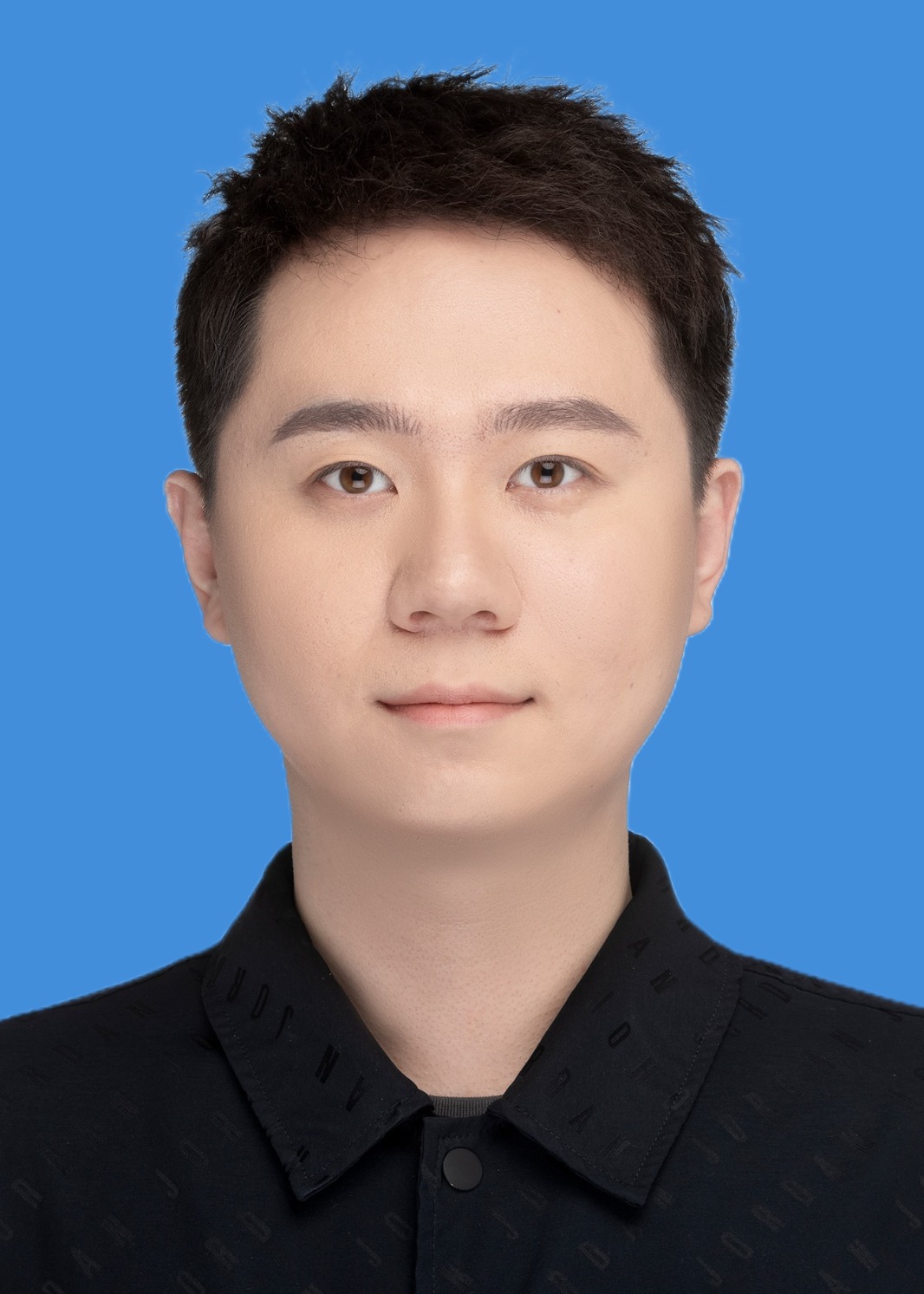}
\end{wrapfigure}\par
\textbf{Lei Chai} (Member, ACM) is currently working toward the Ph.D. degree in software engineering with the School of Computer Science and Engineering, Beihang University. He received his MSc degree from the School of informatics at Xiamen University, China. His research interests include crowdsourcing quality control and human-in-the-loop AI.

\vspace{10pt}

\begin{wrapfigure}{l}{25mm}
\includegraphics[width=1in,height=1.25in,clip,keepaspectratio]{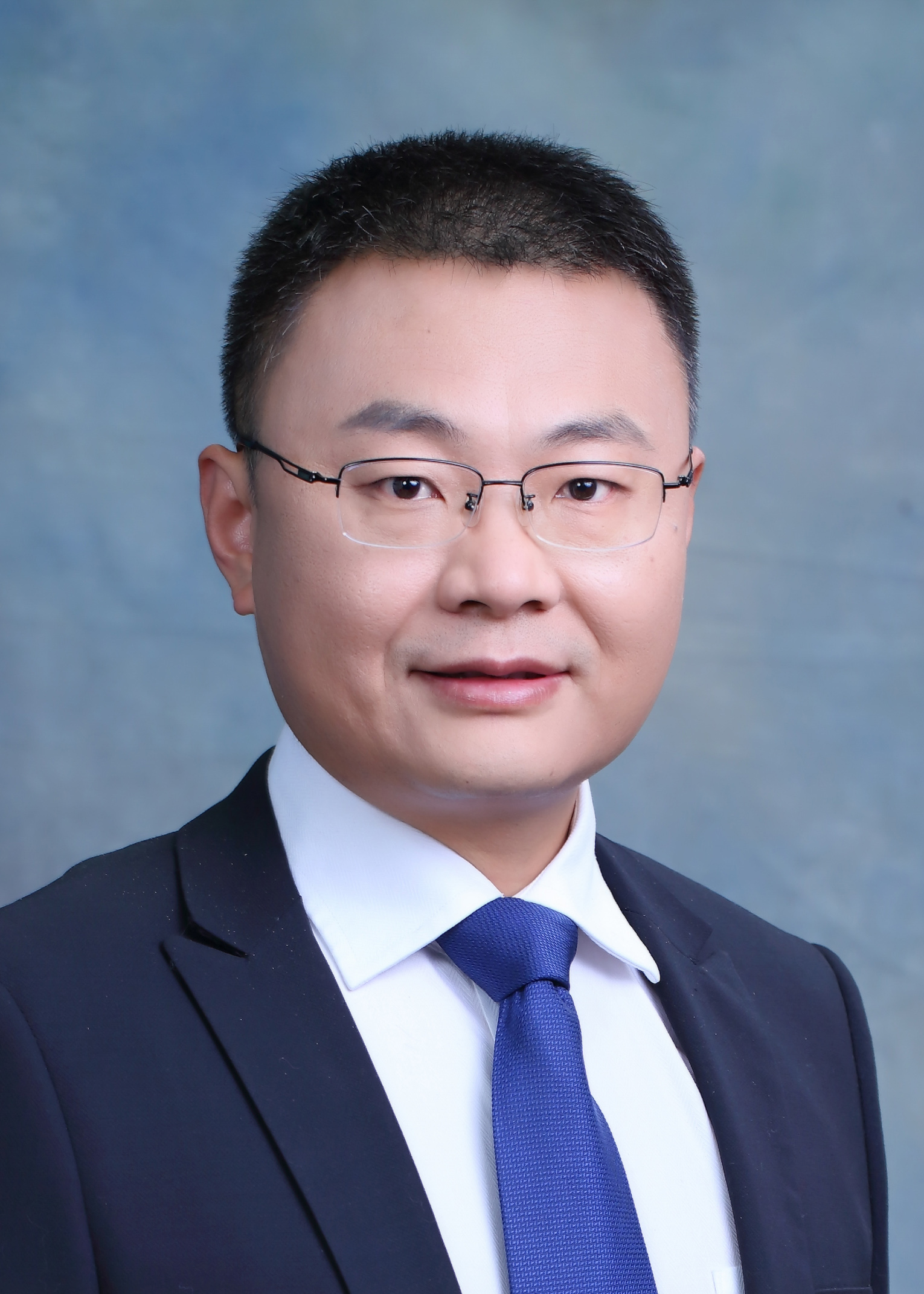}
\end{wrapfigure}\par
\textbf{Hailong Sun} (Member, IEEE) received the B.S. degree in computer science from Beijing Jiaotong University, Beijing, China, in 2001 and the Ph.D. degree in computer software and theory from Beihang University, Beijing, in 2008. His recent research interests include crowd intelligence, intelligent software engineering and distributed system.

\vspace{10pt}

\begin{wrapfigure}{l}{25mm}
\includegraphics[width=1in,height=1.25in,clip,keepaspectratio]{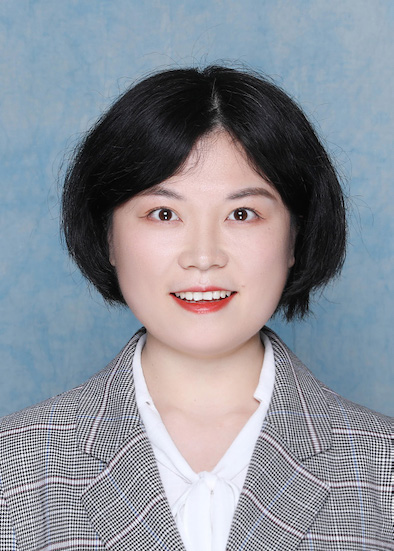}
\end{wrapfigure}\par
\textbf{Jing Zhang} received her Ph.D. degree from University of Wollongong, Australia, in 2019. She is now an associate professor in School of Software, Beihang University, Beijing, China. Her recent research interests include machine learning and computer vision, with the special interests in transfer learning, multi-modality learning, generative models, and their applications in data-efficient computer vision.
\end{document}